\documentclass[pdflatex,sn-mathphys-num]{sn-jnl}

\usepackage{graphicx}
\usepackage{multirow}
\usepackage{amsmath,amssymb,amsfonts}
\usepackage{amsthm}
\usepackage{mathrsfs}
\usepackage[title]{appendix}
\usepackage{xcolor}
\usepackage{textcomp}
\usepackage{manyfoot}
\usepackage{booktabs}
\usepackage{listings}
\usepackage{xspace}
\usepackage{siunitx}
\usepackage{tikz}			
\usepackage{pgfplots}
\usepackage[capitalise]{cleveref}
\usepackage[acronym,shortcuts,nonumberlist]{glossaries}
\usepackage{xurl}
\usepackage{multirow}
\usepackage{booktabs}
\usepackage{float}
\usepackage{subfig}
\usepackage{colortbl}

\DeclareSIUnit{\dB}{\decibel} 

\makeglossaries

\newacronym{HIDVAS}{HIDVAS}{Hearing Instrument Dataset in Various Acoustical Scenarios}

\newacronym{SAL}{SAL}{SONORA Audio Laboratory}
\newacronym{EAL}{EAL}{ExpORL Audio Laboratory}
\newacronym{BTE}{BTE}{behind-the-ear}
\newacronym{ITE}{ITE}{in-the-ear}
\newacronym{RIC}{RIC}{receiver-in-canal}

\newacronym{PCB}{PCB}{printed circuit board}

\newacronym{DRR}{DRR}{direct-to-reverberant ratio}
\newacronym{SPL}{SPL}{sound pressure level}
\newacronym{LUFS}{LUFS}{loudness units full scale}
\newacronym{SER}{SER}{signal-to-error ratio}
\newacronym{RMSE}{RMSE}{root-mean-square error}
\newacronym{RMS}{RMS}{root-mean-square}
\newacronym{SNR}{SNR}{signal-to-noise ratio}
\newacronym{LDR}{LDR}{linear-to-distortion ratio}
\newacronym{NESD}{NESD}{normalised Euclidean system distance}

\newacronym{CSTR}{CSTR}{Centre for Speech Technology Research}
\newacronym{VCTK}{VCTK}{Voice Cloning Toolkit}
\newacronym{LISTf}{LISTf}{Leuven intelligibility sentences test spoken by a female speaker}
\newacronym{LISTm}{LISTm}{Leuven intelligibility sentences test spoken by a male speaker}
\newacronym{SSN}{SSN}{speech shaped noise}
\newacronym{LIST}{LIST}{Leuven intelligibility sentences test}
\newacronym{VCTKf}{VCTKf}{VCTK corpus spoken by female speakers}
\newacronym{VCTKm}{VCTKm}{VCTK corpus spoken by male speakers}

\definecolor{colDHM}{rgb}{0.00000,0.44700,0.74100}
\definecolor{colBTEM}{rgb}{0.85000,0.32500,0.09800}
\definecolor{colXM}{rgb}{0.92900,0.69400,0.12500}
\definecolor{colRIC}{rgb}{0.49400,0.18400,0.55600}
\definecolor{colXS}{rgb}{0.46600,0.67400,0.18800}

\definecolor{RT1}{RGB}{161,218,180 }
\definecolor{RT2}{RGB}{65,182,196}
\definecolor{RT3}{RGB}{44,127,184}
\definecolor{RT4}{RGB}{37,52,148}

\definecolor{windowColor}{RGB}{0, 102, 204} 
\definecolor{wardrobeColor}{RGB}{165, 120, 80} 
\definecolor{curtainColor}{rgb}{1.0, 0.65, 0.79}

\definecolor{silenceColor}{rgb}{0.00000,0.44700,0.74100}
\definecolor{WavColor}{rgb}{0.46600,0.67400,0.18800}

\newcommand{\ie}{i.\,e.,\xspace}

\newcommand{\eg}{e.\,g.,\xspace}

\renewcommand{\Vec}{\mathbf}

\newcommand{\cl}{\texttt{CL}\xspace}

\newcommand{\spk}[2]{\texttt{l#1#2}}

\newcommand{\mic}[2]{\texttt{m#1#2}}

\DeclareSIUnit[]{\SPLunit}{\ensuremath{\,\text{SPL}}}
\DeclareSIUnit[]{\dBSPL}{\dB \SPLunit}

\DeclareSIUnit[]{\SPLequnit}{\ensuremath{\,\text{SPLeq}}}
\DeclareSIUnit[]{\dBSPLeq}{\dB \SPLequnit}

\DeclareSIUnit[]{\fullscale}{\ensuremath{\,\text{FS}}}
\DeclareSIUnit[]{\dBFS}{\dB \fullscale}

\newcommand{\RT}{\ensuremath{\text{T}_{30}}}

\begin{document}

\title[Article Title]{HIDVAS: A Hearing Instrument Dataset in Various Acoustical Scenarios for Algorithm Evaluation and Training}

\author*[1]{\fnm{Arnout} \sur{Roebben}}\email{arnout.roebben@kuleuven.be}
\equalcont{These authors contributed equally to this work.}

\author[1]{\fnm{Giuliano} \sur{Bernardi}}\email{giuliano.bernardi@kuleuven.be}
\equalcont{These authors contributed equally to this work.}

\author[2]{\fnm{Jan} \sur{Wouters}}\email{jan.wouters@kuleuven.be}

\author[1]{\fnm{Toon} \sur{van Waterschoot}}\email{toon.vanwaterschoot@kuleuven.be}

\author[1]{\fnm{Marc} \sur{Moonen}}\email{marc.moonen@kuleuven.be}

\affil[1]{\orgdiv{Department of Electrical Engineering, ESAT-STADIUS}, \orgname{KU Leuven}, \orgaddress{\street{Kasteelpark Arenberg 10}, \city{Leuven}, \postcode{3001}, \country{Belgium}}}

\affil[2]{\orgdiv{Department of Neurosciences, ExpORL}, \orgname{KU Leuven}, \orgaddress{\street{Onderwijs en Navorsing 2 (O\&N2), Herestraat 49}, \city{Leuven}, \postcode{3000}, \country{Belgium}}}

\abstract{To evaluate the performance of audio signal processing algorithms and to train data-driven algorithms, e.g., as applied in hearing instruments, either simulated or recorded data can be used. While large batches of simulated data can be generated using mathematical models, recorded data provide a more adequate representation of real-life scenarios. Therefore, in this paper, the \ac{HIDVAS} is introduced. This dataset consists of both impulse responses and audio recordings using eight external loudspeakers, two external microphones, and a dummy head. On this dummy head \ac{BTE} hearing instrument shells with two microphones per shell are mounted, and in the dummy head's ears \ac{RIC} hearing instrument loudspeakers are inserted. The dummy head also contains microphones located at its eardrum. The impulse responses have been computed from a swept-sine recording for each microphone-loudspeaker pair, and the audio recordings have been obtained by playing back audio (male and female speech, speech shaped noise, singing voice, stringed instrument, wind instrument, and percussion instrument) through each individual loudspeaker and recording simultaneously using all microphones. These recordings have been repeated for four hearing instrument domes (open, semi-open, closed, and no-\ac{RIC}) in three reverberation conditions in one room ($\RT=\SI{0.09}{\second}$, $\RT=\SI{0.47}{\second}$, and $\RT=\SI{0.73}{\second}$), and in one reverberation condition in a different room ($\RT=\SI{1.48}{\second}$). The usage of the dataset as a `hearing instrument in a box' is exemplified with three example use cases. To this end, the ipsilateral feedback paths, the ratio of direct-to-reverberant sound energy in an external microphone close to the sound source versus in the \ac{BTE} hearing instrument microphones, and the leakage from external sound sources to the eardrum are studied. The feedback and leakage for the open and semi-open domes are similar and substantially larger than for the closed dome. The ratio of the direct-to-reverberant sound energy in the \ac{BTE} hearing instrument microphones is lower than in the external microphone, and this difference increases with the reverberation time. The dataset is available at \cite{roebben_dataset,roebben_dataset2}.
}


\maketitle
\glsresetall

\glsresetall

\section{Introduction}
\label{sec:introduction}
To evaluate the performance of audio signal processing algorithms and to train data-driven algorithms, e.g., as applied in hearing instruments \cite{kates_digital_2008}, either simulated or recorded data can be used \cite{brinkmann_round_2019,de_sena_modeling_2015}.
As the generation of simulated data does not require any recording equipment, large batches of simulated data can be generated.
Nevertheless when using these simulated data, the effects of room geometry and absorption are modelled, often requiring simplifying assumptions to make the generation of the simulated data tractable, e.g., assuming the room to be a perfect rectangular `shoebox' \cite{brinkmann_round_2019,de_sena_modeling_2015}.
While the process of obtaining recorded data can be time-consuming, it allows to obtain adequate representations of real-life scenarios with non-regular room geometries and varying absorption properties.
Consequently, recorded data allow to obtain a more adequate representation of real-life scenarios to characterise an algorithm's performance or to train a data-driven algorithm. 

The generation of such recorded data tailored to acoustic hearing instruments commonly employs human listeners or a dummy head, i.e., a head and torso with ears on which hearing instruments can be mounted. Dummy heads offer advantages such as repeatability and the usage of standardised components, while human listeners offer added realism and allow to quantify intersubject variability \cite{denk_hearpiece_2020}.
Datasets that utilise such a dummy head or human listener for hearing instrument applications differ in 1) the loudspeaker configuration, 2) the microphone configuration, 3) the nature of the recorded material, and 4) the recording room conditions.
First, regarding the loudspeaker configuration, loudspeakers can be mounted on an earpiece and inserted into the dummy head's or human listener's ears replicating hearing instrument loudspeakers (commonly referred to as receivers) \cite{sankowsky_on_2015,hettler_acoustic_2024}.
Alternatively, external loudspeakers can be used replicating external sound sources \cite{fejgin_brudex_2023,dietzen_myriad_2023,delebecque_binaurec_2023,woods_real-world_2015,jeub_binaural_2009,kayser_database_2009,maazaoui_romeo-hrtf_nodate,algazi_cipic_2001,noauthor_3d_nodate,lamba_dummy_2024,thiemann_multiple_nodate,sass_comparison_2010,moore_otimp_2019,oreinos_measurement_2013}.
Both loudspeaker configurations can also be used simultaneously \cite{denk_adapting_2018,denk_hearpiece_2020,spriet_combined_2007}.
Second, regarding the microphone configuration, microphones can be mounted on the earpiece replicating hearing instrument microphones \cite{jeub_binaural_2009,spriet_combined_2007,delebecque_binaurec_2023}, microphones can be inserted into the dummy head's or human listener's ear canal to record the signal close to the eardrum \cite{algazi_cipic_2001,noauthor_3d_nodate,lamba_dummy_2024}, or external microphones can be used.
Different microphone configurations can also be used simultaneously, such as considering both microphones on the earpiece and close to the eardrum \cite{woods_real-world_2015,kayser_database_2009,denk_adapting_2018,denk_hearpiece_2020,hettler_acoustic_2024,thiemann_multiple_nodate,sass_comparison_2010,moore_otimp_2019,sankowsky_on_2015,oreinos_measurement_2013}, or considering all microphone configurations simultaneously \cite{fejgin_brudex_2023,dietzen_myriad_2023,maazaoui_romeo-hrtf_nodate}.
Third, regarding the nature of the recorded material, impulse responses can be captured, representing a linear model between each considered loudspeaker-microphone pair \cite{jeub_binaural_2009,kayser_database_2009,spriet_combined_2007,maazaoui_romeo-hrtf_nodate,algazi_cipic_2001,noauthor_3d_nodate,lamba_dummy_2024,thiemann_multiple_nodate,sass_comparison_2010,moore_otimp_2019,oreinos_measurement_2013,sankowsky_on_2015,denk_adapting_2018,denk_hearpiece_2020,hettler_acoustic_2024}.
While this linear model is commonly assumed in algorithm design, e.g., \cite{kates_digital_2008,spriet_combined_2007}, and provides an adequate description of the system, it nevertheless fails to fully capture the non-linearities in the playback and recording equipment, e.g., in the microphones and loudspeakers.
Consequently, audio can also be directly recorded, or both impulse responses and audio can be recorded \cite{fejgin_brudex_2023,dietzen_myriad_2023,delebecque_binaurec_2023,woods_real-world_2015}.
Finally, regarding the recording room conditions, the recording procedure can either be performed in one \cite{delebecque_binaurec_2023,woods_real-world_2015,denk_hearpiece_2020,denk_adapting_2018,hettler_acoustic_2024,spriet_combined_2007,maazaoui_romeo-hrtf_nodate,algazi_cipic_2001,noauthor_3d_nodate,lamba_dummy_2024,thiemann_multiple_nodate,sass_comparison_2010,moore_otimp_2019,sankowsky_on_2015,oreinos_measurement_2013} or multiple conditions \cite{fejgin_brudex_2023,dietzen_myriad_2023,jeub_binaural_2009,kayser_database_2009} thereby, e.g., varying the reverberation time. To this end, the reverberation time can be varied by considering different rooms \cite{jeub_binaural_2009, dietzen_myriad_2023,kayser_database_2009}, or by varying the absorption properties within a room \cite{fejgin_brudex_2023}. To remove the room-dependent effects, anechoic rooms can be considered \cite{kayser_database_2009,denk_adapting_2018,denk_hearpiece_2020,maazaoui_romeo-hrtf_nodate,lamba_dummy_2024,thiemann_multiple_nodate,sass_comparison_2010,oreinos_measurement_2013,noauthor_3d_nodate}.

In this paper, a dummy head is equipped with \ac{BTE} hearing instrument shells, devices mounted behind the ear of the dummy head that contain the hearing instrument microphones, as well as \ac{RIC} hearing instrument loudspeakers, devices where the loudspeaker signal is transmitted electrically to a loudspeaker in the ear. This configuration allows all aforementioned degrees of freedom to be considered simultaneously
\begin{enumerate}
\item by utilising both external loudspeakers replicating external sound sources, and \ac{RIC} hearing instrument loudspeakers mounted on the dummy head's earpiece, replicating the hearing instrument loudspeakers, with four different domes (open, semi-open, closed, and no-\ac{RIC}); 
\item by utilising microphones on the dummy head's earpiece replicating \ac{BTE} hearing instrument microphones (using which software directional microphones could be configured), as well as utilising microphones at the eardrum, and external microphones;
\item by both computing impulse responses from swept-sine recordings between each microphone-loudspeaker pair and recording audio (male and female speech, speech shaped noise, singing voice, stringed instrument, wind instrument, and percussion instrument); 
\item by replicating the recording procedure in four different reverberation times ($\RT=\SI{0.09}{\second}$, $\RT=\SI{0.47}{\second}$, $\RT=\SI{0.73}{\second}$, and $\RT=\SI{1.48}{\second}$). 
\end{enumerate}
The resulting dataset will be referred to as the \ac{HIDVAS}.

As all degrees of freedom are explored simultaneously and all loudspeaker-microphone pairs are considered, \eg also considering the contralateral feedback paths between hearing instrument loudspeakers and microphones, the dataset facilitates its usage as a `hearing instrument in a box'.
To further demonstrate this, in this paper, three example use cases are explored, thereby also motivating the different components (e.g., the multitude of domes, the use of external microphones) included in the dataset. A feedback use case explores the effect of the dome type and reverberation time on the feedback paths between \ac{RIC} hearing instrument loudspeakers and \ac{BTE} hearing instrument microphones, an assisted listening device use case explores the \ac{DRR} both in the external microphone close to the external sound source and in the \ac{BTE} hearing instrument microphones, and a leakage use case explores the propagation of sound from an external source to the eardrum for different domes.

In \cite{denk_hearpiece_2020}, a dataset of impulse responses with a dummy head and different human listeners has been recorded in an anechoic chamber, thereby also considering microphones at the earpiece and eardrum, loudspeakers at the earpiece, and external loudspeakers.
This setup also facilitates the `hearing instrument in a box' application.
Nevertheless, the dataset as presented here is complementary as it also includes external microphones, provides direct recordings of audio signals next to impulse responses, and considers multiple reverberation times.
Additionally, while in \cite{denk_hearpiece_2020} an \ac{ITE} hearing instrument was used, thereby studying devices where both the microphones and loudspeakers are part of the same rigid structure to be inserted into the ear canal, in the present dataset \ac{BTE} hearing instrument microphones are used together with \ac{RIC} hearing instrument loudspeakers. 

Finally, while in \cite{denk_hearpiece_2020} the effect of the hardware is removed, as the primary focus in \cite{denk_hearpiece_2020} is to study the properties of the acoustic paths, in the present dataset the hardware effects are preserved in the recordings. Indeed, the intended goal of the presented dataset is to provide realistic data for algorithm evaluation and training, not to provide and study acoustic impulse responses. The inclusion of the hardware adds realism for evaluation and training as the hardware-in-the-loop affects the recordings in a real-life hearing instrument, and as the effect of hardware and acoustics cannot be disentangled in practice. When recording data for this algorithm evaluation and training, the relative differences between the microphones should be preserved, such that a recording procedure is adhered to to achieve a comparable recording level across microphones. Other recently recorded and documented datasets also incorporate such hardware effects and follow a similar procedure, e.g., \cite{dietzen_myriad_2023,kayser_database_2009,fejgin_brudex_2023,damiano2025trajectorirdatabaseroomacoustic}, and have been used for algorithm evaluation and training, e.g., \cite{li_icassp_2026}.

The paper is structured as follows.
In \cref{sec:setup_description}, the recording rooms, and microphone and loudspeaker configuration are described.
In \cref{sec:recording_chain}, the audio equipment is discussed, and in \cref{sec:recorded_signals}, the recorded impulse responses and audio signals are described. In \cref{sec:analysis}, the dataset is analysed in terms of reverberation time, and the error between the linear impulse response model and the recorded audio. Finally, in \cref{sec:example_use_case} the dataset is further motivated in terms of the three example use cases.

\section{Setup}
\label{sec:setup_description}

\subsection{Recording Rooms}
\label{subsec:room}
\begin{figure}[t]
    \centering
   \includegraphics[width=\columnwidth]{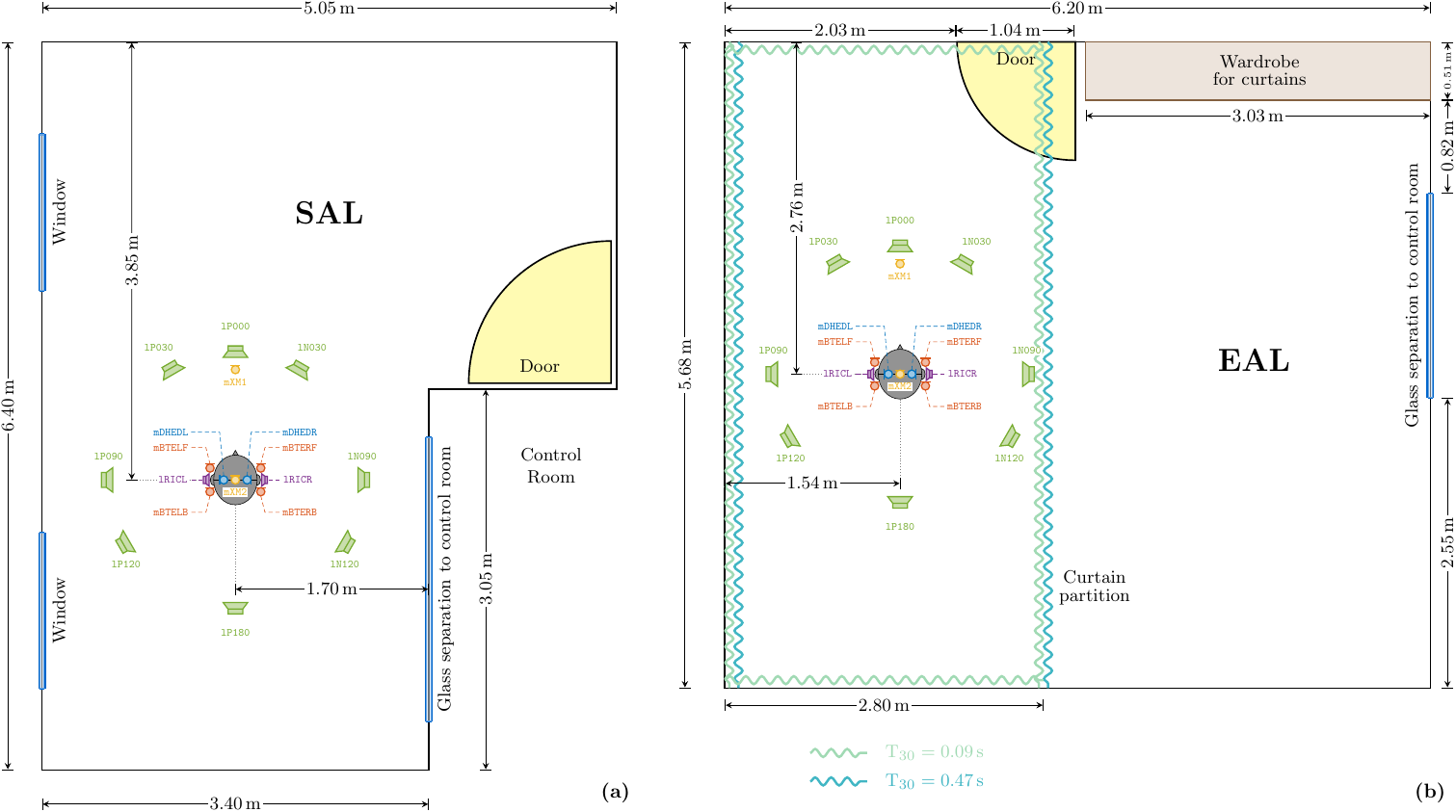}
    \caption{Floor plans of the \ac{SAL} (a) room and \ac{EAL} (b) room, on which the positions of the microphones and loudspeakers are indicated. For the floor plan of the \ac{EAL}, the curtains used for the two lowest reverberation times, \ie\ $\color{RT1}\RT=\SI{0.09}{\second}$ and $\color{RT2}\RT=\SI{0.47}{\second}$, are schematically indicated. The curtain covering the ceiling for $\color{RT1}\RT=\SI{0.09}{\second}$ has been omitted from the floor plan.}
    \label{fig:setup:SAL-EAL-floor-plan}
\end{figure}

\subsubsection{\acf{SAL}} \label{subsubsec:SAL}
The floor plan of the \acf{SAL} \cite{sonora-lab}, located at the Department of Electrical Engineering (ESAT) of KU Leuven (Heverlee, Belgium), is shown in \cref{fig:setup:SAL-EAL-floor-plan}a.
The floor plan of this \SI{3.75}{\meter}-tall room is consistent with the description, schematics, and figures provided in \cite{dietzen_myriad_2023}, where further characteristics of the room have been detailed.
Due to the storage of additional material and equipment within the room, the reverberation time is $\RT = \SI{1.48}{\second}$, estimated as described in section \ref{sec:analysis}, which differs from $\text{T}_{20} = \SI{2.1}{\second}$ as reported in \cite{dietzen_myriad_2023}.

\subsubsection{\acf{EAL}} \label{subsubsec:EAL}
\begin{figure}[t]
    \centering
    \includegraphics[width=0.65\columnwidth]{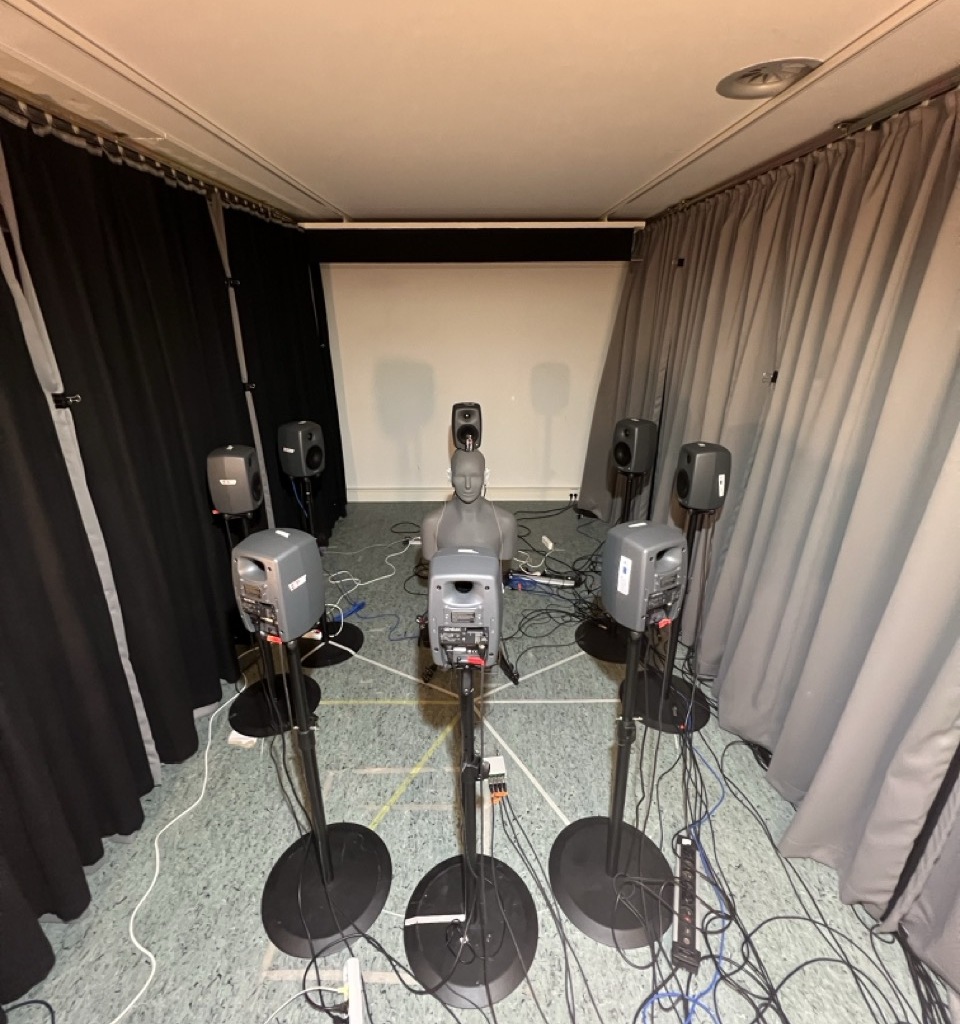}
    \caption{Picture of the setup in the \acf{EAL} leading to \RT=\SI{0.47}{\second}.}
    \label{fig:setup:setup-EAL-pic}
\end{figure}

The floor plan of the \acf{EAL}, located at the Department of Neuroscience (ExpORL) of KU Leuven (Leuven, Belgium), is shown in \cref{fig:setup:SAL-EAL-floor-plan}b.
The reverberation time of this \SI{2.59}{\meter}-tall room can be varied by covering the walls and ceiling using curtains. Additionally, the room can be split in two using a curtain partition.
As schematically indicated in \cref{fig:setup:SAL-EAL-floor-plan}b, the setup was positioned in the left side of the room when facing the door; as also apparent from a picture of the setup in \cref{fig:setup:setup-EAL-pic}.
Three different curtain configurations were used, resulting in different $\RT$-values, which were estimated as described in Section \ref{sec:analysis}:
\begin{itemize}[nosep]
    \item $\RT=\SI{0.09}{\second}$: The curtain partition was used to split the room in two, and all walls in the left side of the room were covered using the curtains as shown in \cref{fig:setup:SAL-EAL-floor-plan}b. Additionally, the ceiling was covered using a curtain. 
    \item $\RT=\SI{0.47}{\second}$: The curtain partition was used to split the room in two.
    The \SI{5.68}{\meter}-long sides of the rectangular left side of the room were covered using the curtains, while the \SI{2.80}{\meter}-long sides were left uncovered as shown in \cref{fig:setup:SAL-EAL-floor-plan}b. The ceiling was left uncovered as well. 
    \item $\RT=\SI{0.73}{\second}$: The curtain partition was left open, and no curtains covered any of the walls or ceiling.  
\end{itemize}

\subsection{Microphone and Loudspeaker Configuration}
\label{subsec:mic_loudspeaker_config}
\begin{figure}[t]
    \centering
    \includegraphics[width=\columnwidth]{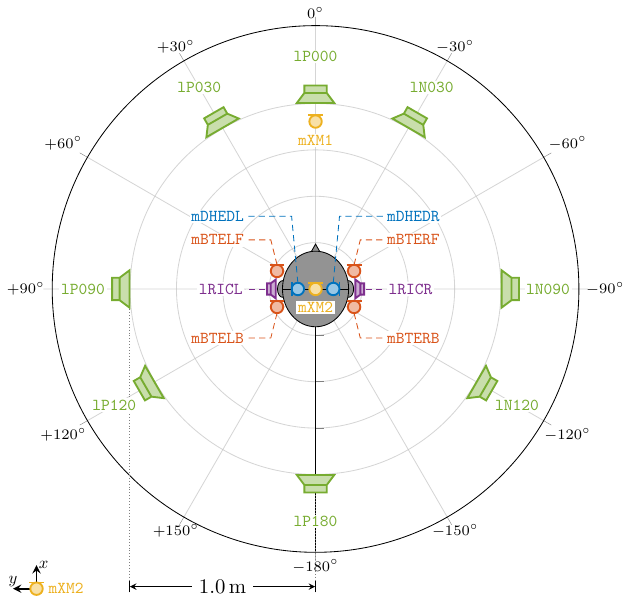}
    \caption{Plan view of the microphone and the loudspeaker configuration.
    The coordinate system on the bottom left indicates the convention used for the coordinates given in the \texttt{coordinates.csv} file that comes with the dataset. The naming convention of the loudspeakers, indicated with \texttt{l} suffix, and of the microphones, indicated with \texttt{m} suffix, is detailed in \cref{table:abbreviations}.}
    \label{fig:setup:micspk-scheme}
\end{figure}

\cref{fig:setup:micspk-scheme} shows the microphone and loudspeaker configuration as positioned both in the \ac{SAL} and \ac{EAL}, and \cref{table:abbreviations} contains the accompanying descriptions of the microphone and loudspeaker naming convention. The centre of the Cortex MK II artificial head and torso is considered as the origin in this configuration. 
This Cortex MK II artificial head and torso is a dummy head with cavities to model the ear canal and on which different models of the outer ear can be mounted.
At the end of the cavity on each side of the head, the dummy head contains a microphone (\mic{DHED}{L} and \mic{DHED}{R}) to record signals at the eardrum. These microphones are MTG condenser microphones, and were located at a height of \SI{1.2}{\meter} in accordance with the height of the dummy head and correspondingly in accordance with the height of a person sitting on a chair.

\begin{figure}[t]
    \centering
    \includegraphics[width=\columnwidth]{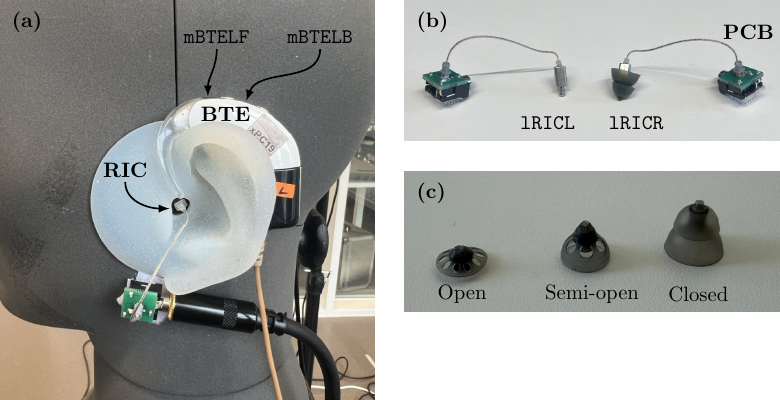}
    \caption{(a) Closeup of the \acf{RIC} hearing instrument loudspeaker and \acf{BTE} hearing instrument microphones placed on the left ear of the dummy head.
    (b) Closeup of the left \ac{RIC} hearing instrument loudspeaker and right \ac{RIC} hearing instrument loudspeaker (without and with dome) mounted on their respective \acp{PCB}.
    (c) Closeup of the \ac{RIC} hearing instrument loudspeaker domes (open, semi-open, closed).}
    \label{fig:setup:setup-cl-RIC-fits}
\end{figure}

On top of the dummy head ears, type SP15 \ac{BTE} hearing instruments provided by Cochlear Ltd.
were mounted.
Each of these \ac{BTE} hearing instruments contains two microphones (\mic{BTE}{LF}, \mic{BTE}{LB}, \mic{BTE}{RF}, and \mic{BTE}{RB}).
Two Sonion Steel Series XP \ac{RIC} hearing instrument loudspeakers, as shown in \cref{fig:setup:setup-cl-RIC-fits}b, were inserted in the ear canal to serve as hearing instrument loudspeakers (\spk{RIC}{L} and \spk{RIC}{R}). 
The insertion of the \ac{RIC} hearing instrument loudspeakers was done in a way that the posterior tip of the \ac{RIC} hearing instrument loudspeaker was aligned with the entrance of the ear canal of the dummy head.
To keep the \ac{RIC} hearing instrument loudspeakers in place, three different domes with variable occlusion were used as shown in \cref{fig:setup:setup-cl-RIC-fits}c. Ranging from least to most occluding, the Phonak open dome large 4.0, Phonak open dome large 3.0, and Phonak power dome large 3.0 were used. These domes will be further referred to in terms of occlusion as open, semi-open, and closed respectively. A condition without insertion of the \ac{RIC} hearing instrument loudspeakers was also considered, and will be referred to as the no-\ac{RIC} condition. A closeup of the left \ac{BTE} hearing instrument microphones and \ac{RIC} hearing instrument loudspeaker mounted on the dummy head is shown in \cref{fig:setup:setup-cl-RIC-fits}a.

Eight Genelec 8030CP loudspeakers, used to model external sound sources, were positioned on loudspeaker stands on a circular grid with a \SI{1}{\meter} radius at the angular positions \SI{0}{\degree}, \SI{\pm30}{\degree}, \SI{\pm90}{\degree}, \SI{\pm120}{\degree}, and \SI{180}{\degree}. Herein, \SI{0}{\degree} is defined as the frontal direction of the dummy head, \ie\ the direction the dummy head is facing.
The corresponding loudspeaker positions are labeled as \spk{P}{000}, \spk{P}{030}, \spk{N}{030}, \spk{P}{090}, \spk{N}{090}, \spk{P}{120}, \spk{N}{120}, and \spk{P}{180}. Herein, the \texttt{P} and \texttt{N} naming convention refers to positive and negative angles respectively, and the three succeeding digits refer to the angle at which the loudspeaker is positioned. As these loudspeakers are external to the dummy head configuration, they will be referred to as external loudspeakers. The external loudspeakers were located at a height of \SI{1.1}{\meter}.

Finally, two AKG CK32 microphones, external to the dummy head configuration and hence referred to as external microphones, were considered. These microphones are consequently referred to as \mic{XM}{1} and \mic{XM}{2}.
To this end, \mic{XM}{1} was positioned close to the frontal loudspeaker at the same height as this loudspeaker at a distance of \SI{0.17}{\meter} to model the use of assisted listening devices. This \mic{XM}{1} microphone was oriented horizontally towards the \spk{P}{000} loudspeaker.
The microphone \mic{XM}{2} was positioned on top of the dummy head above its centre position to allow for microphone recordings less influenced by the head shadow effect of the dummy head. This \mic{XM}{2} microphone was oriented vertically towards the ceiling.

For further details about loudspeaker and microphone positions, all the 3D coordinates with respect to the centre of the dummy head are provided in a \texttt{coordinates.csv} file in the dataset.

\begin{table}
\arrayrulecolor{black}
\begin{tabular}{ l l}
\hline
\textbf{Abbreviation} & \textbf{Description} \\
\cmidrule(lr){2-2}
\mic{BTE}{LF} & Microphone behind-the-ear left front\\
\mic{BTE}{LB} & Microphone behind-the-ear left back\\
\mic{BTE}{RF} & Microphone behind-the-ear right front\\
\mic{BTE}{RB} & Microphone behind-the-ear right back\\
\mic{DHED}{L} & Microphone dummy head ear drum left\\
\mic{DHED}{R} & Microphone dummy head ear drum right\\
\mic{XM}{$x$} & Microphone external microphone $x$\\
\cmidrule(lr){2-2}
\spk{RIC}{L} & Loudspeaker receiver-in-canal left\\
\spk{RIC}{R} & Loudspeaker receiver-in-canal right\\
\spk{P}{$xxx$} & Loudspeaker positioned at $xxx\si{\degree}$\\
\spk{N}{$xxx$} & Loudspeaker positioned at -$xxx\si{\degree}$\\
\hline
\end{tabular}
\caption{Abbreviations and accompanying descriptions of microphones and loudspeakers as shown in \cref{fig:setup:micspk-scheme}.}
\label{table:abbreviations}
\end{table}

\section{Recording Chain}
\label{sec:recording_chain}

\subsection{Hardware} \label{subsec:hardware}
\begin{figure}[h]
    \centering
    \includegraphics[width=\columnwidth]{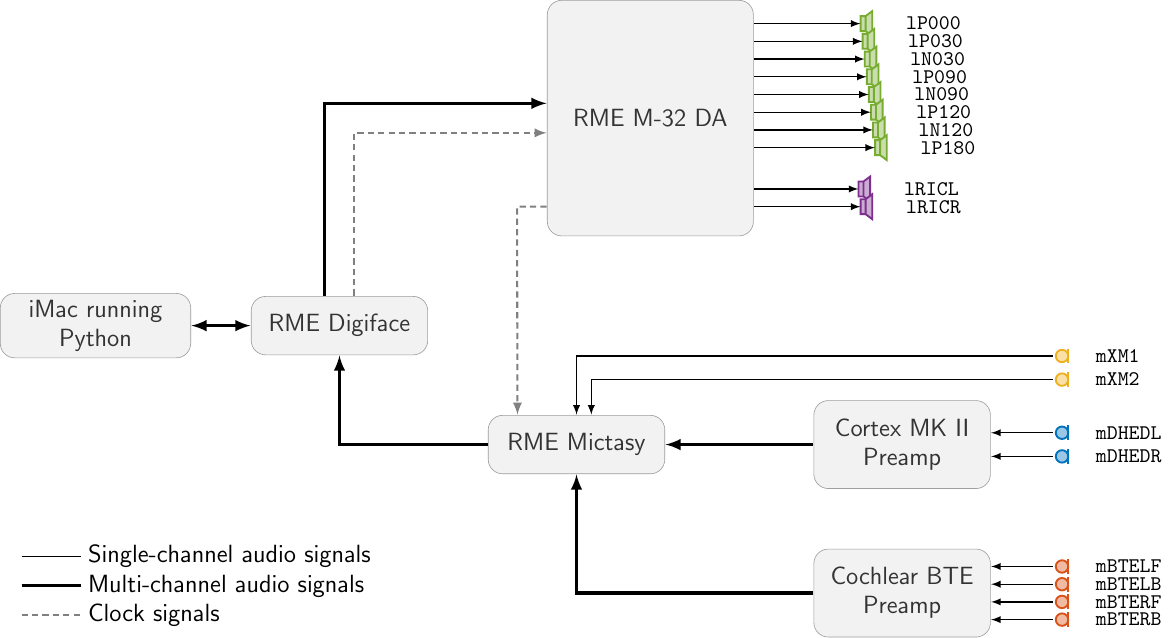}
    \caption{Recording chain indicating the different components used for the acquisition and playback of audio signals.}
    \label{fig:recording_chain}
\end{figure}

The recording chain is shown in \cref{fig:recording_chain}, and was identical for the \ac{SAL} and the \ac{EAL}. An iMac running Python 3.10 was used to control the audio equipment for playback and recording as a digital audio workstation. The signals were fed to the playback chain, and fetched from the acquisition chain via an RME Digiface audio interface connected to the iMac via USB. The recorded microphone signals were stored on the iMac at a sampling frequency $f_s=\SI{48}{\kilo\hertz}$ using a bit depth of \SI{16}{\bit}.

For the playback chain, the RME Digiface audio interface was connected to an RME M-32 DA multi-channel digital-to-analog converter. 
The transfer of the playback signals from the RME Digiface to the RME M-32 DA was done using the ADAT protocol. 
The output of the RME M-32 DA was then connected to the eight Genelec 8030CP external loudspeakers and the two Sonion Steel Series XP \ac{RIC} hearing instrument loudspeakers.

For the acquisition chain, the RME Digiface audio interface was connected to an RME Micstasy preamplifier/converter. 
The transfer of the recorded signals from the  RME Micstasy to the RME Digiface was done using the ADAT protocol.
The AKG CK32 external microphones were connected directly to the RME Micstasy, while the \ac{BTE} hearing instrument microphones and dummy head ear drum microphones first passed through dedicated preamplifiers provided by Cochlear Ltd. and Cortex Instruments respectively. The Cortex Instruments preamplifier was the original preamplifier accompanying the Cortex MK II artificial head and torso.  

As for the clock signals, the RME Digiface was used as the primary clock, to which the RME M-32 DA was connected as the secondary clock via an ADAT cable. 
In turn, the RME Micstasy was connected to the RME M-32 DA to receive the clock signal via a BNC cable. 

\subsection{Gain Adjustment}
\label{subsec:preparation_calibration}
\begin{figure}[h]
    \centering
    \includegraphics[width=0.6\columnwidth]{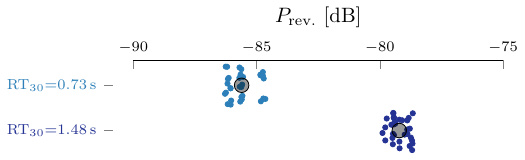}    
    \caption{The power in the reverberant tail for the impulse responses for each of the external loudspeakers to \mic{XM}{2} are in agreement, confirming the validity of the loudspeaker gain adjustment. The black dots represent the average across loudspeakers.}
    \label{fig:calibration:Rpower}
\end{figure}

Prior to the measurements in the \ac{SAL} and the \ac{EAL} rooms, the software and hardware gains of the loudspeakers and microphones were determined and the hardware delay was compensated for. This procedure aims at achieving comparable recording levels across the microphones in order to maintain comparable relative differences between microphones, which is important for the intended goal of algorithm evaluation and training \cite{dietzen_myriad_2023,kayser_database_2009}. Other hardware effects were not compensated for as the dataset functions as a `hearing instrument in a box', where the hardware-in-the-loop does affect the recordings, and to allow for a plug-and-play usage of the dataset. Indeed, the goal of the dataset is not to provide acoustic impulse responses, which would require hardware to be switched in for algorithm evaluation and training. The following procedure was used:

\begin{enumerate}
    \item \textbf{External Loudspeaker Gain Adjustment:} The software gains of the external loudspeakers, positioned on a circle at \SI{1}{\meter} distance from the centre of the circular grid (\cref{fig:setup:micspk-scheme}), were adjusted such that each loudspeaker achieved an equal pressure at the centre of this  grid.
    To this end, a B\&K Type 2250 \ac{SPL} meter was placed at the centre of this circular grid at a \SI{1.2}{\meter} height, \ie\ at the same height as the dummy head ear drum microphones and at the same distance from each of the external loudspeakers, by temporarily removing the dummy head. The \ac{SPL} meter was tilted upwards.
    The software gain values in the control software were adjusted to ensure an equal output level at the \ac{SPL} meter of \SI{85}{\dBSPL} for each of the individual external loudspeakers playing back \ac{SSN}, resulting in the software gain values used throughout the recordings. These gain values are reported in \cite{roebben_dataset,roebben_dataset2}. To verify these gain values, as per \cite{damiano2025trajectorirdatabaseroomacoustic}, the power in the reverberant tail $P_{\text{rev.}}$ for each of these loudspeakers to the \mic{XM}{2} microphone was calculated. For a fixed room, this reverberant power only depends on the power of the loudspeaker signal, and on hence the associated software gain \cite{naylor_speech_2010}. As shown in \cref{fig:calibration:Rpower}, the reverberant power for each of these impulse responses is in agreement, with a maximum deviation between any two loudspeakers of \SI{1.6}{\decibel}.

    \item \textbf{Microphone Gain Adjustment:} The microphone gains were subsequently adjusted such that a diffuse sound field resulted in the same overall recording level in each of the microphones. To this end, after loudspeaker gain adjustment, the dummy head itself, containing the dummy head ear drum microphones and \ac{BTE} hearing instrument, and the external microphones were put in place, while the \ac{RIC} hearing instrument loudspeakers were still excluded. A diffuse sound field was created by positioning a Genelec 8030CP loudspeaker and a Genelec 8020DPM loudspeaker in opposite corners of the room and tilting those loudspeakers outwards to face the corners, by positioning the \ac{SPL} meter at \SI{1}{\meter} height between the dummy head and the \mic{XM}{1} microphone, by adjusting the software gain values in the control software such that each of the additional Genelec 8030CP and 8020DPM loudspeakers separately playing back \ac{SSN} produced \SI{70}{\dBSPL} at the \ac{SPL} meter, and finally by playing back the same \ac{SSN} through both loudspeakers simultaneously.
    The sound field was verified to be sufficiently diffuse in the measurement area by placing the \ac{SPL} meter at different positions in the room where the measurement microphones would be placed and checking that the \ac{RMS} power of the recorded signals were within a \SI{1.5}{\dBSPL} tolerance. Using this diffuse sound field, the gains on the RME Micstasy associated to the \ac{BTE} hearing instrument and dummy head microphones, and to the external microphones were adjusted to ensure that the recordings were at the same level (within a \SI{1.5}{\dB} tolerance) for this diffuse sound field. Additionally, the operating gain was chosen roughly midway the dynamic range of the RME Micstasy for the software gain values of the external loudspeakers as determined in the previous step in order to maximise the dynamic range of the recordings. These gain values are reported in \cite{roebben_dataset,roebben_dataset2}.

    \item \textbf{\ac{RIC} hearing instrument Loudspeaker Gain Adjustment:} The software gains of the \spk{RIC}{R} (\spk{RIC}{L}) loudspeaker were adjusted to achieve an equal pressure at the \mic{DHED}{R} (\mic{DHED}{L}) microphone as with the \spk{N}{090} (\spk{P}{090}) loudspeaker.
    To this end, \ac{SSN} was played back from the \spk{N}{090} (\spk{P}{090}) loudspeaker and the microphone signal at the corresponding \mic{DHED}{R} (\mic{DHED}{L}) microphone was recorded without the \spk{RIC}{R} (\spk{RIC}{L}) loudspeaker inserted. 
    After inserting the \spk{RIC}{R} (\spk{RIC}{L}) loudspeaker with the \cl dome, the microphone signal was recorded again with playback from this \spk{RIC}{R} (\spk{RIC}{L}) loudspeaker. 
    The gain values in the control software for the \spk{RIC}{R} (\spk{RIC}{L}) loudspeaker were then adjusted so that the \ac{RMS} power of the microphone signals were within \SI{0.5}{\dB} tolerance to achieve an equivalent sound level at the eardrum.

    \item \textbf{Hardware Latency Compensation:} The latency of the recording and playback equipment was measured using a calibration pulse signal. In the impulse responses and recorded audio this latency is compensated for, such that the direct component in the recordings matches with the travel time of a sound wave from the source to the microphone. 
\end{enumerate}

\section{Recorded Signals}
\label{sec:recorded_signals}
The dataset contains the impulse responses for all microphone-loudspeaker pairs, and audio (male and female speech, speech shaped noise, singing voice, stringed instrument, wind instrument, and percussion instrument) played back by each individual loudspeaker and recorded by all microphones simultaneously. 
The recording swept-sines and computation of the resulting impulse responses is described in \cref{subsec:sweeps} and the recording of the audio in \cref{subsec:audio}. 

In total, the dataset contains \SI{14592}{files} of raw audio data, consisting of \SI{1216}{files} of recorded sweeps for impulse response measurements and \SI{13376}{files} of other recorded audio, \SI{1216}{files} of computed impulse responses, and \SI{11}{files} of audio sources, for a total of \SI{15819}{files}.
The raw audio data amounts to \SI{156.39}{\hour} of recordings, with an additional \SI{1.01}{\hour} of computed IR data and \SI{0.12}{\hour} of source audio data, for a total of \SI{157.53}{\hour} of audio data. The dataset occupies \SI{50.34}{\giga\byte} of storage for the raw files, \SI{0.65}{\giga\byte} for the computed impulse response files, and \SI{0.04}{\giga\byte} for the source files, totaling \SI{51.03}{\giga\byte}. 

\subsection{Impulse Responses}
\label{subsec:sweeps}
To determine the impulse response for a microphone-loudspeaker pair, the exponential sine sweep method was used, wherein a sine with exponentially increasing frequency excites the system and then the recorded signal is deconvolved with the original swept-sine to obtain the impulse response \cite{farina_2000_02,farina_2007_05}. 

\begin{figure}[h]
    \centering
    \includegraphics[width=1.0\columnwidth]{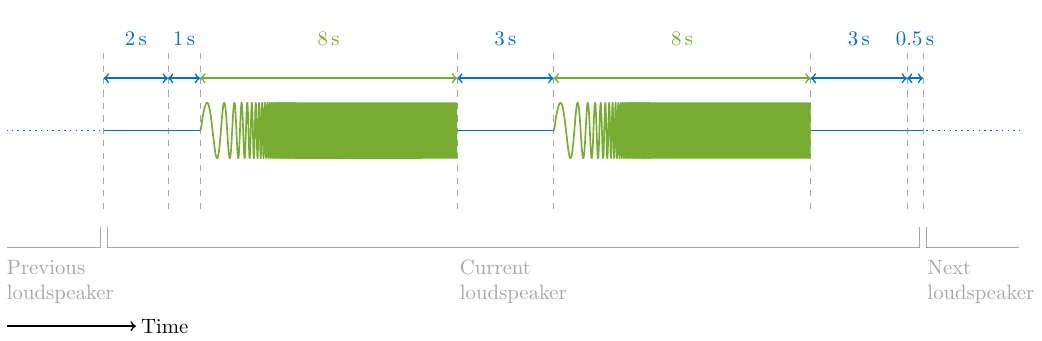}   
    \caption{Presentation structure of the exponential sine sweeps. The silence interleaving the sweeps avoids intersweep interference and allows for latency compensation.}
    \label{fig:sweep}
\end{figure}

For each microphone-loudspeaker pair, two repetitions of an \SI{8}{\second}-long swept-sine followed by \SI{3}{\second} of silence were played as shown schematically in \cref{fig:sweep}. This \SI{3}{\second} of silence was chosen to be larger than the largest reverberation time of the room ($\RT=\SI{1.48}{\second}$) to avoid intersweep interference Additionally, \SI{1}{\second} of silence was added before the first sweep and \SI{0.5}{\second} of silence was added after the second sweep to allow for latency compensation in case the second sweep was used for deconvolution. Finally, an additional silence of \SI{2}{\second} was inserted between the measurements of any two consecutive microphone-loudspeaker pairs; this was not strictly necessary for the impulse response measurements, but it was retained  for consistency with the audio recordings. This presentation structure of the sweeps is schematically illustrated in Fig. \ref{fig:sweep}. The swept-sines were designed to excite a frequency region ranging from \SI{1}{\hertz} to \SI{24}{\kilo\hertz} according to \cite{farina_2000_02,farina_2007_05}, although other swept-sines exist, e.g., \cite{novak2015synchronized}. Before playback, the gain of the swept-sines was tuned to achieve the same \ac{RMS} level as the \ac{SSN} from the \ac{VCTK} corpus (cfr. \cref{subsec:audio}).

The deconvolution was performed using the first of the recorded swept-sines unless the first recorded swept-sine was noisy (\eg due to impulsive noise) as identified based on visual inspection. 
In that case, the second swept-sine was used for the deconvolution\footnote{The second swept-sine was only used for 2 of the recordings.}. Averaging the impulse responses from both recordings was explicitly avoided as per \cite{farina_2007_05}.
The hardware delay in the resulting impulse responses was compensated for as detailed in \cref{subsec:preparation_calibration} such that the direct component of the impulse responses matches with the travel time of sound waves from the loudspeakers to the microphones. The resulting impulse responses were saved at a bit depth of \SI{32}{\bit}. 
Other hardware effects were not compensated for as the dataset functions as a `hearing instrument in a box', where the hardware-in-the-loop indeed does affect the recordings.

\subsection{Audio}
\label{subsec:audio}
Although the impulse responses adequately model the input-output relationship between the loudspeaker and microphone signals, as will be shown in \cref{sec:analysis}, there are still non-linear effects in the loudspeakers and microphones which cannot be fully captured by the impulse responses. 
Consequently, audio is recorded as well, across a wide range of speech and music classes (male and female speech, speech shaped noise, singing voice, stringed instrument, wind instrument, and percussion instrument) by playing back the audio through each individual loudspeaker, one at a time, and recording through all microphones simultaneously.

Four types of speech materials were included in the dataset, i.e., Flemish and English speech as spoken by both female and male speakers.
The Flemish female speech materials were sampled from the corpus of the \ac{LISTf} \cite{vanwieringen2008listf}, and the Flemish male speech materials were sampled from the corpus of the \ac{LISTm} \cite{jansen2014listm}. 
Both corpora are typically used for audiological testing \cite{vanwieringen2008listf,jansen2014listm}. 
The English female and male materials were sampled from the corpus of the \ac{CSTR} \ac{VCTK} \cite{yamagishi2019cstrvctk}, commonly employed in the evaluation of speech enhancement algorithms \cite{dubey2022icassp2022deepnoise}.

More specifically, for the \ac{LISTf} material, one speech file of duration \SI{54.9}{\second} was created by concatenating the ten sentences from \texttt{list01} (\texttt{001.wav} - \texttt{010.wav}) and the first five sentences from \texttt{list02} (\texttt{001.wav} - \texttt{005.wav}).
Each of the sentences was separated by \SI{2.5}{\second} of silence to avoid intersentence interference, as such interference would disturb an audiological test administered using the recorded material.\footnote{This value was chosen to be between \SI{1.5}{times} and \SI{2}{times} the largest reverberation time of the room ($\RT=\SI{1.48}{\second}$).} 
Similarly, for the \ac{LISTm} material, one speech file of duration \SI{65.0}{\second} was created by concatenating the ten sentences from \texttt{list01} (\texttt{001.wav} - \texttt{010.wav}) and the first five sentences from \texttt{list02} (\texttt{001.wav} - \texttt{005.wav}). 
Again, \SI{2.5}{\second} of silence was inserted between each of the sentences to avoid intersentence interference.
Finally, for the \ac{VCTK} female and male materials, one female and one male speech file of duration \SI{30.6}{\second} and \SI{30.7}{\second} respectively, was created by concatenating randomly sampled sentences from female and male speakers.
Contrary to the \ac{LISTf} and \ac{LISTm} material, no intersentence silence was added between the sentences of the \ac{VCTK} materials, as these \ac{VCTK} materials were included to mimic continuous natural speech. 

Three types of \ac{SSN}, of \SI{20.0}{\second} each, were included in the dataset, i.e., \ac{SSN} calculated using the \ac{LISTf} corpus, the \ac{LISTm} corpus, and the \ac{VCTK} corpus. These \ac{SSN} materials were created by modifying the spectrum of white noise to match the average spectrum of the respective corpora, and are commonly employed to mask speech signals \cite{francart2011comparison}.

Four types of music materials (singing voice, stringed instrument, wind instrument, and percussion instrument) were included in the dataset in order to cover a range of timbres and instruments classes, e.g., to be used for music processing \cite{roa2025first}. The singing voice was selected from the `Anechoic recordings of Italian opera' corpus \cite{dorazio2020anechoicOperaDataset}, and the other music materials from the `Music for Archimedes' album by Bang \& Olufsen \cite{hansen1991archimedesDataset} (based on \cite{hansen1991archimedes}). 

More specifically, for the singing voice, one music file (\texttt{mu\_Soprano.wav}) of duration \SI{30.8}{\second} was created by selecting a part of the original audio file wherein a lyric soprano performs a portion of the aria `Di tale amor, che dirsi' from \textit{Il Trovatore} by Giuseppe Verdi. 
For the stringed instrument, wind instrument, and percussion instrument \SI{30.8}{\second}, \SI{31.1}{\second}, and \SI{30.4}{\second} files were created by selecting parts of the original cello (\texttt{Track22.wav}), trumpet (\texttt{Track37.wav}), and bongo (\texttt{Track26.wav}) files respectively, on the `Music for Archimedes' album. Fade-out was applied to the selected portion of the source audio when necessary. 

All speech and music materials were recorded in (hemi-)anechoic rooms \cite{vanwieringen2008listf,jansen2014listm,yamagishi2019cstrvctk,dorazio2020anechoicOperaDataset,hansen1991archimedesDataset}. 
Before playback, all the speech and music \texttt{.wav} files were additionally normalised to achieve a target of \SI{-23}{\decibel} on the \ac{LUFS} according to the ITU-R BS.1770-4 standard \cite{ITU-R-BS.1770-4} using the pyloudnorm toolbox \cite{pyloudnorm}. 
During playback, a silence of \SI{2}{\second} was inserted between each of the audio files to avoid interfile interference. 
As for the impulse responses, the hardware delay was compensated for, such that the delay of the audio recordings matches the delay of the impulse responses convolved with the sound sources.

\section{Analysis}
\label{sec:analysis}
In \cref{subsec:rt30}, the $\RT$ estimates of the reverberation times are discussed to characterise the acoustic properties of the rooms, and in \cref{subsec:impulse_response_versus_recorded_audio} the recorded audio signals are compared to the signals obtained by convolving the source audio with the impulse responses to validate the applicability of the impulse response model.

\subsection{\RT}
\label{subsec:rt30}
\begin{figure}[h]
    \centering
    \includegraphics[width=1.0\columnwidth]{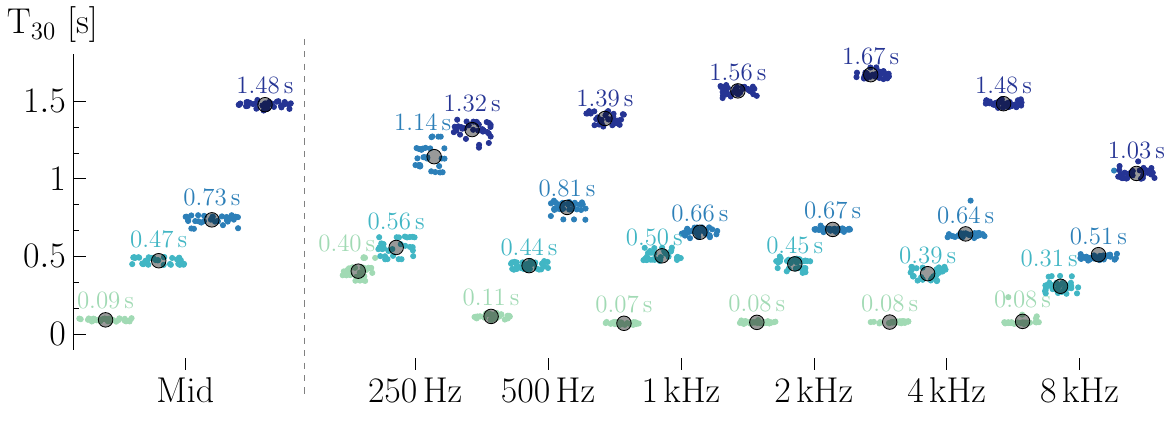} 
    \caption{Estimated $\RT$ averaged across the \SI{500}{\hertz} and \SI{1}{\kilo\hertz} octave bands (mid) and in six individual octave bands, obtained from the impulse responses between the external loudspeakers and the \mic{XM}{2} external microphone. Each impulse response was remeasured four times, as the measurement procedure was repeated for the four dome types and the dome does not affect the reverberation time. 
    Each data point represents the estimate obtained from one impulse response, and the black dots represent the average across loudspeaker positions (and domes). The numeric value of this average is shown as well. The order of the estimates per frequency band and the associated color scheme correspond, from left to right, to $\color{RT1}\RT=\SI{0.09}{\second}$, $\color{RT2}\RT=\SI{0.47}{\second}$, $\color{RT3}\RT=\SI{0.73}{\second}$, and $\color{RT4}\RT=\SI{1.48}{\second}$.}
    \label{fig:analysis:rt30_frequency_analysis}
\end{figure}

\cref{fig:analysis:rt30_frequency_analysis} shows the averaged (\ie mid) and per-octave-band-frequency estimates (\ie\ at 250 Hz, 500 Hz, 1 kHz, 2 kHz, 4 kHz, and 8 kHz) of the $\RT$ estimates obtained from the impulse responses between each external loudspeaker and the \mic{XM}{2} external microphone according to the methodology specified in the ISO 3382-1 standard \cite{ISO-3382-1} and implemented using the IoSR MATLAB Toolbox \cite{iosr_matlab_toolbox}. This procedure was repeated four times, as there are four dome types and these domes do not affect the reverberation time.
The mid $\RT$ estimate has been obtained as the average $\RT$ across the \SI{500}{\hertz} and \SI{1}{\kilo\hertz} octave bands conform the ISO 3382-1 standard \cite{ISO-3382-1}.
As shown by the individual data points in \cref{fig:analysis:rt30_frequency_analysis}, little deviation from the mean $\RT$ is found across different loudspeaker positions (and domes).

\subsection{Impulse Response Versus Recorded Audio}
\label{subsec:impulse_response_versus_recorded_audio}
\begin{figure}[h]
    \centering
    \subfloat[Singing voice with $\RT=\SI{0.09}{\second}$]{
    \includegraphics[width=\columnwidth]{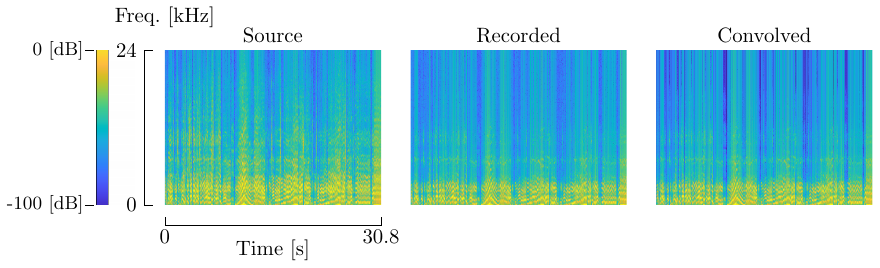} 
    }\vfill 
    \subfloat[\ac{VCTK} male speech $\RT=\SI{1.48}{\second}$]{
    \includegraphics[width=\columnwidth]{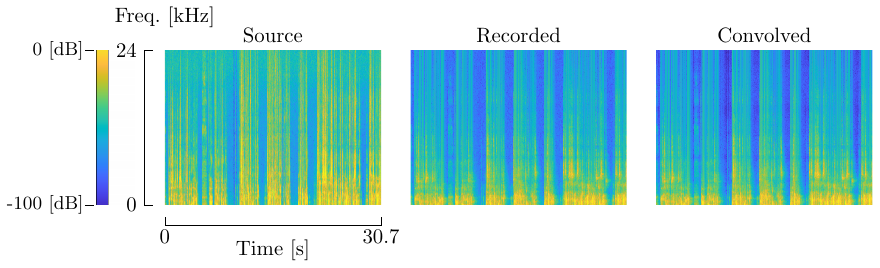}  
    }   
    \caption{Spectrograms of source audio, recorded audio, and source audio convolved with the impulse response between \spk{P}{000} and \mic{XM}{1}. The recorded and convolved audio visually match well. The source audio has been normalised to have the same \acf{RMS} value as the recorded audio.}
    \label{fig:analysis:spectra}
\end{figure}

\begin{figure}[h]
    \centering
    \makebox[\textwidth][c]{
    \subfloat[\ac{SER}]{
    \begin{minipage}{0.49\textwidth}
    \includegraphics[width=\columnwidth]{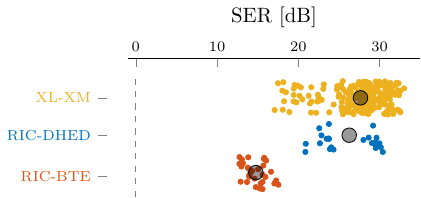}  
   \label{fig:analysis:vctk_error_metrics_with_exclusions:ser}
    \end{minipage} 
    }\hfill 
    \subfloat[\ac{LDR}]{
    \begin{minipage}{0.49\textwidth}
    \includegraphics[width=\columnwidth]{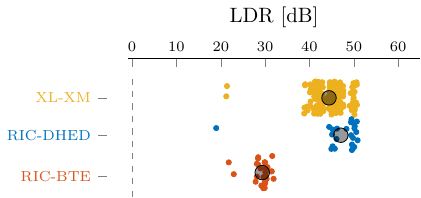}  
   \label{fig:analysis:vctk_error_metrics_with_exclusions:ldr}
   \end{minipage} 
    }    
    } 
    \caption{The \acf{SER} for the \acf{VCTK} male speech audio and the \acf{LDR}. Both metrics are computed between each of the external loudspeakers and the external microphones ({\color{colXM}XL-XM}), between the \spk{RIC}{R} (\spk{RIC}{L}) loudspeaker and the dummy head ear drum microphone at the same side \mic{DHED}{R} (\mic{DHED}{L}) ({\color{colDHM}RIC-DHED}), and between the \spk{RIC}{R} (\spk{RIC}{L}) loudspeaker and the \acf{BTE} hearing instrument microphones at the same side \mic{BTE}{RF} and \mic{BTE}{RB} (\mic{BTE}{{LF}} and \mic{BTE}{LB}) ({\color{colBTEM}RIC-BTE}).
    All domes and reverberation times are considered simultaneously, although the closed dome is excluded from the {\color{colBTEM}RIC-BTE} condition due to the low \acf{SNR} of the feedback signal.
    Individual data points represent one recording.
    The mean across domes and rooms is represented as a black dot.}
    \label{fig:analysis:vctk_error_metrics_with_exclusions}
\end{figure}

\cref{fig:analysis:spectra} shows the spectra of source audio, recorded audio, and the source audio convolved with the impulse response. The spectra of recorded and convolved audio match well, and illustrate the effect of smearing due to the reverberation. In order to further confirm the validity of this linear model of the impulse responses, the \ac{SER} is analysed between the recorded audio signal and the source audio signal convolved with the corresponding impulse response \cite{dietzen_myriad_2023,trainor2004development}.
Denoting the recorded audio signal at time $k$ as $a[k]$, the source audio signal as $s[k]$ with $\Vec{s}[k]=\begin{bmatrix}s[k] & \cdots & s[k-L+1]\end{bmatrix}^\top\in\mathbb{R}^{L\times 1}$, and the impulse response as $\Vec{h}=\begin{bmatrix}h[0] & \cdots & h[L-1]\end{bmatrix}^\top \in\mathbb{R}^{L\times 1}$, the \ac{SER} is defined as
\begin{equation}
    \text{SER} =20\log_{10}\left(\frac{\sqrt{\frac{1}{K}\sum_{k=0}^{K-1}a[k]^2}}{\sqrt{\frac{1}{K}\sum_{k=0}^{K-1}\left(a[k]-\Vec{h}^\top\Vec{s}[k]\right)^2}}\right),
\end{equation}
with $K$ the total length of the audio recording. For additional validation, next to the linear model $\Vec{h}$, the exponential sine sweep also comes with a model $\Vec{h}_n\in\mathbb{R}^{L_n\times1}$ for the non-linear distortion in the system \cite{farina_2000_02,farina_2007_05}. Correspondingly, a \ac{LDR} metric is defined as follows
\begin{equation}
\text{LDR} =20\text{log}_{10}\left(\frac{\sqrt{\frac{1}{L}\sum_{k=0}^{L-1}h[k]^2}}{\sqrt{\frac{1}{L_n}\sum_{k=0}^{L_n-1}h_n[k]^2}}\right).
\end{equation}
\cref{fig:analysis:vctk_error_metrics_with_exclusions:ser} shows the \ac{SER} with the \ac{VCTK} male speech for different loudspeaker-microphone pairs that are recorded at a sufficiently high \ac{SNR}, \ie\ the \ac{SER} between each of the external loudspeakers and both the external microphones, between the \spk{RIC}{R} (\spk{RIC}{L}) loudspeaker and the dummy head ear drum microphone at the same side \mic{DHED}{R} (\mic{DHED}{L}), and between the \spk{RIC}{R} (\spk{RIC}{L}) loudspeaker and the \ac{BTE} hearing instrument microphones at the same side \mic{BTE}{RF} and \mic{BTE}{RB} (\mic{BTE}{LF} and \mic{BTE}{LB}).
All domes and reverberation times are considered simultaneously, yet the closed dome was excluded when considering the impulse response from the \spk{RIC}{R} (\spk{RIC}{L}) loudspeaker to \mic{BTE}{RF} and \mic{BTE}{RB} (\mic{BTE}{LF} and \mic{BTE}{LB}) microphones due the closed dome having low \ac{SNR} (cfr. \cref{subsec:feedback}).

Each of the individual data points exceeds an \ac{SER} of \SI{12}{\decibel}, and each of the mean \ac{SER} values exceeds \SI{14}{\decibel} as shown in \cref{fig:analysis:vctk_error_metrics_with_exclusions:ser}. These mean values are in agreement with \cite{trainor2004development}, where mean values of \SI{12}{\decibel} were found for the SER for similar recordings.

\cref{fig:analysis:vctk_error_metrics_with_exclusions:ldr} shows the \ac{LDR} for the same scenarios. The \ac{LDR} in \cref{fig:analysis:vctk_error_metrics_with_exclusions:ldr} demonstrates the same patterns as the \ac{SER} in \cref{fig:analysis:vctk_error_metrics_with_exclusions:ser}. Consequently, the impulse responses provide an adequate model for the responses from the loudspeakers to the microphones.

\section{Example Use Cases}
\label{sec:example_use_case}
To illustrate the applicability of the dataset, three example use cases motivate the different components included in the dataset. To this end, the following example use cases are studied: a feedback use case exploring the effect of the dome and reverberation time on the feedback path from the \ac{RIC} hearing instrument loudspeaker to the ipsilateral \ac{BTE} hearing instrument microphone (\cref{subsec:feedback}), an assisted listening device use case exploring the \ac{DRR} both in the external microphone close to the external sound source and in the \ac{BTE} hearing instrument microphones (\cref{subsec:assisted_listening}), and a leakage use case exploring the propagation of sound from an external source as played by an external loudspeaker to the dummy head ear drum microphones for different domes (\cref{subsec:anc}). Consequently, the feedback use case motivates the inclusion in the dataset of the \ac{RIC} hearing instrument loudspeakers and different domes; the assisted listening device use case motivates the inclusion of the external microphones; and the leakage use case motivates the inclusion of the different domes.

\subsection{Feedback Use Case}
\label{subsec:feedback}
\begin{figure}[h]
    \centering
    \subfloat[Impulse response (Linear scale)]{
    \includegraphics[width=\columnwidth]{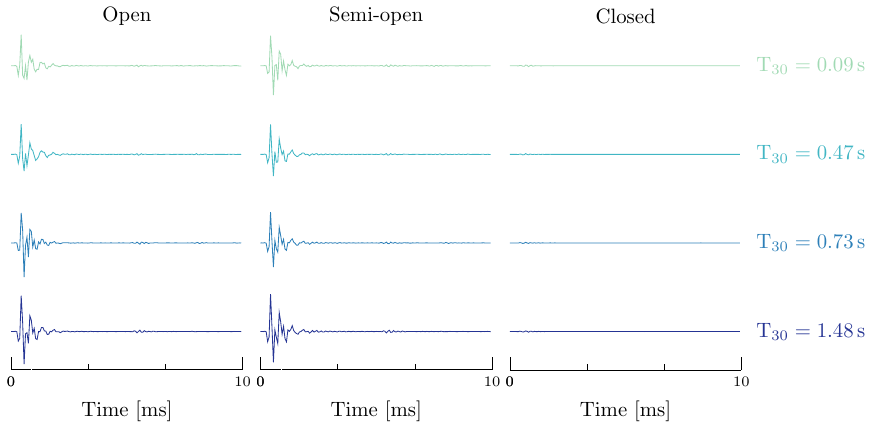}  
    }\vfill 
    \subfloat[{Magnitude spectrum [dB]}]{
    \includegraphics[width=\columnwidth]{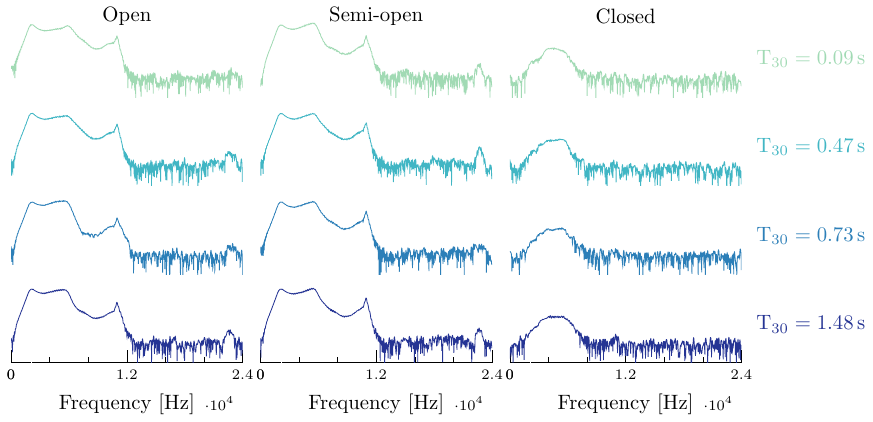}}   
    \caption{Comparison of the ipsilateral feedback paths between \spk{RIC}{R} loudspeaker and \mic{BTE}{RF} microphone for the different reverberation times and domes. All paths are shown with the same magnitude, for which the numeric values are omitted due to this magnitude being of arbitrary unit. Across domes, the open and semi-open feedback paths are similar in magnitude, while the closed feedback paths are smaller in magnitude. Across reverberation times, the feedback paths are similar.}
    \label{fig:analysis:ricr_to_bterf_comparison}
\end{figure}

\cref{fig:analysis:ricr_to_bterf_comparison} shows the impulse responses and magnitude spectra of the feedback paths between the \spk{RIC}{R} loudspeaker and \spk{BTE}{RF}  microphone, across domes and reverberation times. 

Across domes, the open and semi-open domes result in a similar magnitude of the feedback path as the direct path and early reflections have a similar magnitude despite the open dome having more venting (cfr. \cref{fig:setup:setup-cl-RIC-fits}c).
The closed dome, having no venting, blocks a larger proportion of the sound leaking from the ear canal, resulting in a smaller magnitude of the feedback path as shown in \cref{fig:analysis:ricr_to_bterf_comparison}.

Across reverberation times, the overall shape and magnitude of the feedback paths also remains similar for each particular dome due to the path from \spk{RIC}{R} loudspeaker to \mic{BTE}{RF} microphone being short (in the $\SI{}{\centi\meter}$ order of magnitude) and being mainly affected by the dummy head geometry rather than the room itself, indicating only a minor effect of the reverberation time. This is further quantified by the \ac{NESD}, calculated as \cite{lydaki_deep_feedback_2025}
\begin{equation}
    \text{NESD} = 20\log_{10}\left(\frac{\sqrt{\frac{1}{L}\sum_{k=0}^{L-1}{(\Vec{h}_{\text{T},\text{d}}[k]-\Vec{h}_{\SI{0.09}{\second},\text{d}}[k])^2}}}{\sqrt{\frac{1}{L}\sum_{k=0}^{L-1}{\Vec{h}_{\SI{0.09}{\second},\text{d}}[k]}^2}}\right).
\end{equation}
To this end, the \ac{NESD} quantifies the distance between the impulse response for each reverberation time $\text{T}\in\{\SI{0.47}{\second},\SI{0.73}{\second},\SI{1.48}{\second}\}$ and dome $\text{d}\in\{\text{open, semi-open, closed}\}$ with respect to the impulse response recorded with the reverberation time $\SI{0.09}{\second}$ and the same dome $\text{d}$. In \cref{table:NESD}, the \ac{NESD} is shown for the \spk{RIC}{R} - \mic{BTE}{RF} and \spk{RIC}{L} - \mic{BTE}{LF} loudspeaker - microphone combinations. Before calculating the NESD, the impulse responses were time aligned. While the \ac{NESD} generally increases with the reverberation time, the \ac{NESD} remains significantly negative, indicating only a minor effect of the reverberation time.

\begin{table}
\arrayrulecolor{black}
\makebox[\textwidth][c]{
\subfloat[\spk{RIC}{R} - \mic{BTE}{RF}]{
\begin{minipage}{0.45\textwidth}
\begin{tabular}{ l l l l }
\hline
\textbf{\RT} & \textbf{open} &  \textbf{semi-open} & \textbf{closed} \\
\cmidrule(lr){2-4}
\SI{0.47}{\second} & $-12.28$ & $-8.97$ & $-7.78$\\
\SI{0.73}{\second} & $-2.76$ & $-10.54$ & $-6.98$\\
\SI{1.48}{\second} & $-3.93$ & $-11.06$ & $-7.69$\\
\hline
\end{tabular}
\end{minipage}
}

\hfill

\subfloat[\spk{RIC}{L} - \mic{BTE}{LF}]{
\begin{minipage}{0.45\textwidth}
\begin{tabular}{ l l l l }
\hline
\textbf{\RT} & \textbf{open} &  \textbf{semi-open} & \textbf{closed} \\
\cmidrule(lr){2-4}
\SI{0.47}{\second} & $-20.05$ & $-19.24$ & $-8.90$\\
\SI{0.73}{\second} & $-9.46$ & $-9.02$ & $-5.06$\\
\SI{1.48}{\second} & $-9.19$ & $-9.59$ & $-1.72$\\
\hline
\end{tabular}
\end{minipage}
}
}
\caption{\Acf{NESD} between the impulse response recorded with $\RT=\SI{0.09}{\second}$ and the impulse responses with the corresponding dome and loudspeaker - microphone combination at different reverberation times.}
\label{table:NESD}
\end{table}

\subsection{Assisted Listening Device Use Case}
\label{subsec:assisted_listening}
\begin{figure}[h]
    \centering
    \includegraphics[width=0.8\columnwidth]{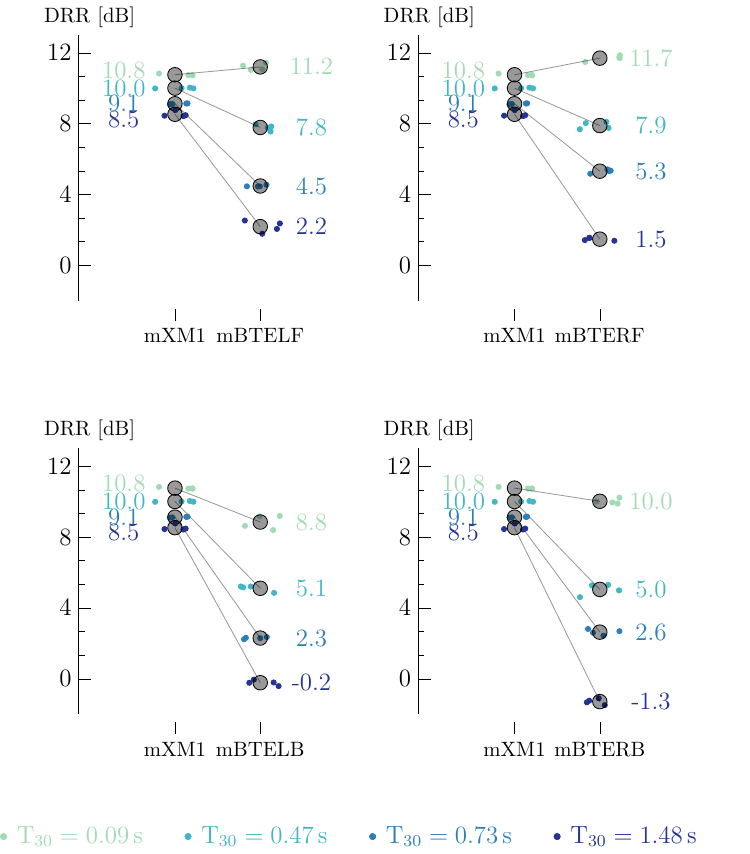}        
    \caption{The \acf{DRR} between the \spk{P}{000} loudspeaker and both the \mic{XM}{1} external microphone and \acf{BTE} hearing instrument microphones and reverberation times. Each \ac{DRR} was remeasured four times, as the measurement procedure was repeated for four dome types and the dome does not affect the \ac{DRR} in the \mic{XM}{1} external microphone or the \ac{BTE} hearing instrument microphones.
    The individual data points represent different domes, the black dots represent the average across domes, and the numbers display the numeric value of this average.
    The \ac{DRR} is larger in the \mic{XM}{1} external microphone, which is closer to the \spk{P}{000} loudspeaker than the \ac{BTE} hearing instrument microphones.
    The \ac{DRR} increase from the \mic{XM}{1} external microphone to the \ac{BTE} hearing instrument microphones is larger as the reverberation time increases.}
    \label{fig:analysis:P000_DRR_consolidated_clouds}
\end{figure}

\cref{fig:analysis:P000_DRR_consolidated_clouds} shows the \ac{DRR} for the impulse responses from the frontal \spk{P}{000} loudspeaker to both the \mic{XM}{1} external microphone close to the loudspeaker (\SI{0.17}{\meter}) and the \ac{BTE} hearing instrument microphones far away from the loudspeaker ($\approx \SI{1}{\meter}$). These \ac{DRR} measurements were repeated four times, as there are four dome types and these domes do not affect the \ac{DRR} in the \mic{XM}{1} external microphone or the \ac{BTE} hearing instrument microphones. 
This \ac{DRR} compares the energy of the direct sound to the energy of the reverberation, and is defined as \cite{naylor_speech_2010}
\begin{equation} \label{eq:drr}
    \text{DRR} = 10\log_{10}\left(\frac{\sum_{k=k_0-\delta}^{k_0+\delta}h[k]^2}{\sum_{k=k_0+\delta+1}^{L-1}h[k]^2}\right).
\end{equation}
Herein, $k_0$ represents the arrival time in samples of the direct sound, $\delta$ defines a temporal region in samples around this arrival time, thereby partially incorporating some early reflections, and $L$ is the length of the impulse response $h[k]$. The \ac{DRR} was calculated using the IoSR MATLAB Toolbox \cite{iosr_matlab_toolbox} with $\frac{\delta}{f_s}=\SI{2.5}{\milli\second}$ and $f_s$ the sampling frequency. Higher \ac{DRR} values have been shown to correlate well with increased speech intelligibility \cite{naylor_speech_2010}. Consequently, assisted listening devices aim at utilising a microphone placed close to the speaker of interest, in order to stream the recordings directly to the hearing instrument \cite{dillon_hearing_2012}.

As shown in \cref{fig:analysis:P000_DRR_consolidated_clouds}, the \ac{DRR} in the \mic{XM}{1} external microphone is larger than the \ac{DRR} in the \ac{BTE} hearing instrument microphones due to the smaller distance from \mic{XM}{1} to \spk{P}{000}.
Indeed, while the energy of the direct sound follows an inverse square law related to the distance between loudspeaker and microphone, the energy of the reverberation is unaffected by the distance \cite{naylor_speech_2010}. The unexpected small \ac{DRR} increase from \mic{XM}{1} to \mic{BTE}{LF} and \mic{BTE}{RF} for \RT=\SI{0.09}{\second} can be attributed to the spread of the estimates and to the usage of a temporal region defined by $\delta$, such that some early reflections are incorporated. The \ac{DRR} is, additionally, larger in the frontal than in the rear \ac{BTE} hearing instrument microphones due to the increased distance from the \spk{P}{000} loudspeaker.
Finally, the difference between the \ac{DRR} in the \mic{XM}{1} external microphone and the \ac{DRR} in the \ac{BTE} hearing instrument microphones increases as the reverberation time increases as the energy of the reverberation increases with the reverberation time \cite{naylor_speech_2010}.
Consequently, the effectiveness of using an external microphone in assisted listening devices is demonstrated, especially when the reverberation time is large.

\subsection{Leakage Use Case}
\label{subsec:anc}
\begin{figure}[h]
    \centering
    \includegraphics[width=0.65\columnwidth]{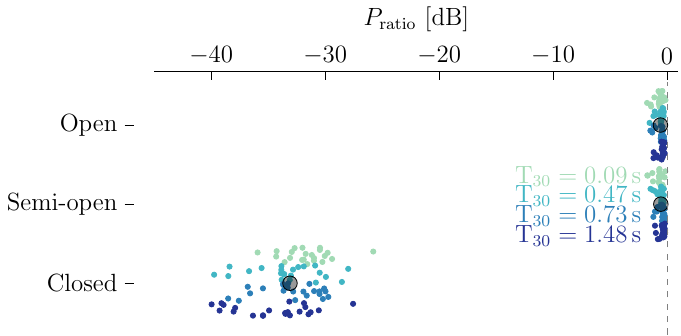}    
    \caption{The ratio of the power of the audio recordings played back by the \spk{P}{000} loudspeaker as measured at the \mic{DHED}{R} (\mic{DHED}{L}) microphone for each dome ($\text{d}\in\{\text{open}, \text{semi-open}, \text{closed}\}$) compared to the power of the audio recordings as measured at the \mic{DHED}{R} (\mic{DHED}{L}) microphone when there is no dome inserted (no-\acf{RIC}).
    The individual data points represent different audio recordings, and the black dots represent the average across audio recordings and reverberation times.
    The open and semi-open domes do not reduce this leakage from the external sound source into the ear canal well, while the closed dome does so.}
    \label{fig:analysis:hearing_aid_to_dh_power_ratio}
\end{figure}

\cref{fig:analysis:hearing_aid_to_dh_power_ratio} shows the power ratio $P_{\text{ratio}}$ of the audio signals recorded in the dummy head ear drum microphones when there is a dome inserted (open, semi-open, and closed) and the audio signals recorded in the dummy head ear drum microphones when there is no dome inserted (no-\ac{RIC}).
This power ratio  $P_{\text{ratio}}$ quantifies the amount of leakage blocked by each of the domes compared to the no-\ac{RIC} case.
Indeed, ideally, only the sound processed by the hearing instrument as played back by the \ac{RIC} hearing instrument loudspeaker should reach the eardrum, while the unprocessed sound from the external sound source should be blocked.
Consequently, this  $P_{\text{ratio}}$ between audio signal $a_\text{d}[k]$ at the \mic{DHED}{R} (\mic{DHED}{L}) microphone for dome $\text{d}\in\{\text{open},\text{semi-open},\text{closed}\}$ and audio signal $a_\text{NR}[k]$ at the \mic{DHED}{R} (\mic{DHED}{L}) microphone for the no-\ac{RIC} case, defined as 
\begin{equation}
    P_{\text{ratio}} = 20\log_{10}\left(\frac{\sqrt{P_\text{d}}}{\sqrt{P_\text{NR}}},\right)
\end{equation}
with $P_\text{d}=\frac{1}{K}\sum_{k=0}^{K-1}a_\text{d}[k]^2$ and $P_\text{NR}=\frac{1}{K}\sum_{k=0}^{K-1}a_\text{NR}[k]^2$, should be as small as possible. This $P_{\text{ratio}}$ is assessed using the frontal loudspeaker 1P000.

Similar to the observations in Section \ref{subsec:feedback}, the open and semi-open domes do not significantly block the external sound source leaking into the ear canal due to their venting, such that $P_{\text{ratio}}$ in \cref{fig:analysis:hearing_aid_to_dh_power_ratio} is close to \SI{0}{\decibel} and similar for both dome types.
On the contrary, the closed dome has no venting and blocks the sound leakage from the \spk{P}{000} loudspeaker into the ear canal, leading to a mean $P_{\text{ratio}}$ of \SI{-33.1}{\decibel}.

\glsresetall

\section{Conclusion}
\label{sec:conclusion}
\glsresetall
In this paper, the \ac{HIDVAS} has been introduced. This dataset consists of both impulse responses and audio recordings (male and female speech, speech shaped noise, singing voice, stringed instrument, wind instrument, and percussion instrument), which were recorded using eight external loudspeakers, two external microphones, and a dummy head on which \ac{BTE} hearing instrument shells were mounted with two microphones per shell, and in which \ac{RIC} hearing instrument loudspeakers were inserted. 
The dummy head also has microphones, located at the ear drum, which were used to record microphone signals at the eardrum of the dummy head. These recordings were repeated for four hearing instrument domes (open, semi-open, closed, and no-\ac{RIC}) in three reverberation conditions ($\RT=\SI{0.09}{\second}$, $\RT=\SI{0.47}{\second}$, and $\RT=\SI{0.73}{\second}$) in the \ac{EAL}, and in one reverberation condition ($\RT=\SI{1.44}{\second}$) in the \ac{SAL}.

The domes affected the feedback paths and leakage, with the closed dome blocking more of the feedback and leakage than the open an semi-open domes. These open and semi-open domes, additionally, resulted in similar feedback paths and leakage. The reverberation time only had a minor effect on the feedback paths. The external microphones, placed close to the external loudspeakers resulted in an increase in \ac{DRR}, which is correlated with increased speech intelligibility, thereby motivating the use of the dataset for assisted listening device use cases as well.

The dataset is available at \cite{roebben_dataset,roebben_dataset2}, and the code for playback, recording, and analysis is available at \cite{roebben_github}.

\clearpage

\backmatter
\section*{List of abbreviations}
\printglossary[type=\acronymtype]
\printglossary
\textit{BTE} behind-the-ear.\\

\noindent\textit{CSTR} Centre for Speech Technology Research.\\

\noindent\textit{DRR} direct-to-reverberant ratio.\\

\noindent\textit{EAL} ExpORL Audio Laboratory.\\

\noindent\textit{HIDVAS} Hearing Instrument Dataset in Various Acoustical Scenarios.\\

\noindent\textit{ITE} in-the-ear.\\

\noindent\textit{LDR} linear-to-distortion ratio.\\
\textit{LISTf} Leuven intelligibility sentences test spoken by a female speaker.\\
\textit{LISTm} Leuven intelligibility sentences test spoken by a male speaker.\\
\textit{LUFS} loudness units full scale.\\

\noindent\textit{NESD} normalised Euclidean system distance.\\

\noindent\textit{PCB} printed circuit board.\\

\noindent\textit{RIC} receiver-in-canal.\\
\textit{RMS} root-mean-square.\\

\noindent\textit{SAL} SONORA Audio Laboratory.\\
\textit{SER} signal-to-error ratio.\\
\textit{SNR} signal-to-noise ratio.\\
\textit{SPL} sound pressure level.\\
\textit{SSN} speech shaped noise.\\

\noindent\textit{VCTK} Voice Cloning Toolkit.

\section*{Declarations}

\bmhead{Availability of data and materials}
The \ac{HIDVAS} dataset is openly accessible at \cite{roebben_dataset}, except for the inclusion of the \ac{LISTf} and \ac{LISTm} material and accompanying \ac{SSN} recordings due to license restrictions. At \cite{roebben_dataset}, the dataset is available either in full, thus including both audio recordings and impulse responses, or in an economic version containing only impulse responses. The dataset including the \ac{LISTf} and \ac{LISTm} material and accompanying \ac{SSN} recordings is available at \cite{roebben_dataset2} together with the information on how to comply with the \ac{LISTf} and \ac{LISTm} material licenses. 

The code used for playback and recording is available in a GitHub repository \cite{roebben_github}. In the same GitHub repository \cite{roebben_github}, the code used to generate the results in \cref{subsec:preparation_calibration} (\cref{fig:calibration:Rpower}), in \cref{sec:analysis} (\cref{fig:analysis:rt30_frequency_analysis}-\cref{fig:analysis:vctk_error_metrics_with_exclusions}), and in \cref{sec:example_use_case} (\cref{table:NESD} and \cref{fig:analysis:ricr_to_bterf_comparison}-\cref{fig:analysis:hearing_aid_to_dh_power_ratio}) is available as well.

\bmhead{Competing interests}
The authors declare that they have no competing interests.

\bmhead{Funding}
This research was carried out at the ESAT Laboratory of KU Leuven in the frame of Research Council KU Leuven C14-21-0075 ”A holistic approach to the design of integrated and distributed digital signal processing algorithms for audio and speech communication devices”, and Aspirant Grant 11PDH24N (for AR) from the Research Foundation - Flanders (FWO), and at the ExpORL laboratories of the Dept. Of Neurosciences, with support of the EU-project MSCA-DN-CherISH-101120054 "Cochlear implants and spatial hearing: Enabling access to the next dimension of hearing" (JW) and VLAIO-O\&O-PreENI-HBC.2024.0184 "Precision Electrode Neural Interface for Cochlear Implants" (JW).

\bmhead{Authors' contributions}
AR, GB, JW, TVW, and MM designed the setup of the dataset and the recording chain. AR and GB carried out the recordings, and post-processed and analysed the data. JW, TVW, and MM contributed to the analysis of the data. AR and GB wrote the draft of the manuscript. All authors read and revised
the final manuscript.

\bmhead{Acknowledgements}
The authors would also like to thank Jan P.\ Koopman from Sonion for providing the \ac{RIC} hearing instrument loudspeakers, Lieselot Van Deun from UZ Leuven for providing the different hearing instrument domes, and Thomas Dietzen from KU Leuven for assisting with the preparation of the recordings. Additionally, the authors would like to thank the Central Services for Electronics and Mechanics of the Department of Electrical Engineering (ESAT) of KU Leuven, including Joren Van Cleynenbreugel, for facilitating the manufacturing of the \acp{PCB} on which the \ac{RIC} hearing instrument loudspeakers were mounted.

\bibliography{dhha-bibliography.bib}

@misc{roebben_dataset,
	author = {Roebben, Arnout and Bernardi, Giuliano and Wouters, Jan and {van Waterschoot}, Toon and Moonen, Marc},
	howpublished = {Zenodo},
	title = {{HIDVAS}: A Hearing Instrument Dataset in Various Acoustical Scenarios for algorithm evaluation and training},
	month=jun,
	year = {2026},
	note = {https://doi.org/10.5281/zenodo.15773862}
}

@misc{roebben_dataset2,
	author = {Roebben, Arnout and Bernardi, Giuliano and Wouters, Jan and {van Waterschoot}, Toon and Moonen, Marc},
	howpublished = {Zenodo},
	title = {{HIDVAS}: A Hearing Instrument Dataset in Various Acoustical Scenarios for algorithm evaluation and training [Full version including license-restricted material]},
	month=jun,
	year = {2026},
	note = {https://doi.org/10.5281/zenodo.15774050}
}

@misc{roebben_github,
  author = {Roebben, Arnout and Bernardi, Giuliano},
  title = {HIDVAS: A Hearing Instrument Dataset in Various Acoustical Scenarios for algorithm evaluation and training},
  year = {2025},
  howpublished = {GitHub repository},
  url = {https://github.com/Arnout-Roebben/HIDVAS}
}

@misc{sonora-lab,
  author = {{van Waterschoot}, Toon},
  title = {{KU} Leuven {ESAT-STADIUS} {Audio} {Research} {Labs}},
  year = {2022},
  url = {https://lirias.kuleuven.be/3940173}
}

@article{dietzen_myriad_2023,
	title = {{MYRiAD}: {A} {Multi}-{Array} {Room} {Acoustic} {Database}},
	volume = {2023},
	language = {en},
	number = {17},
	journal = {EURASIP J. Audio, Speech, Music Process.},
	author = {Dietzen, Thomas and Ali, Randall and Taseska, Maja and {van Waterschoot}, Toon},
	month = apr,
	year = {2023},
	pages = {17},
}

@article{hansen1991archimedes,
  title={Making recordings for simulation tests in the Archimedes project},
  author={Hansen, Villy and Munch, Gert},
  journal={J. Audio Eng. Soc.},
  volume={39},
  number={10},
  pages={768--774},
  year={1991},
  publisher={Audio Engineering Society}
}

@misc{hansen1991archimedesDataset,
	author = {Bang and Olufsen},
	howpublished = {CD B\&O 101},
	title = {Music for Archimedes},
	year = {1992}
}

@misc{dorazio2020anechoicOperaDataset,
	author = {D'Orazio, D.},
	howpublished = {Zenodo},
	title = {Anechoic recordings of Italian opera},
	month=jan,
	year = {2020},
	doi = {10.5281/zenodo.3628247}
}

@misc{yamagishi2019cstrvctk,
  title={CSTR VCTK Corpus: English multi-speaker corpus for CSTR Voice Cloning Toolkit (version 0.92)},
  author={Yamagishi, Junichi and Veaux, Christophe and MacDonald, Kirsten and others},
  journal={University of Edinburgh. The Centre for Speech Technology Research (CSTR)},
  pages={271--350},
  year={2019},
  month=nov,
  url = {https://datashare.ed.ac.uk/handle/10283/3443}
}

@misc{dubey2022icassp2022deepnoise,
      title={ICASSP 2022 Deep Noise Suppression Challenge}, 
      author={Harishchandra Dubey and Vishak Gopal and Ross Cutler and Ashkan Aazami and Sergiy Matusevych and Sebastian Braun and Sefik Emre Eskimez and Manthan Thakker and Takuya Yoshioka and Hannes Gamper and Robert Aichner},
      year={2022},
      eprint={2202.13288},
      archivePrefix={arXiv},
      primaryClass={eess.AS},
      url={https://arxiv.org/abs/2202.13288}, 
}

@article{francart2011comparison,
  title={Comparison of fluctuating maskers for speech recognition tests},
  author={Francart, Tom and Van Wieringen, Astrid and Wouters, Jan},
  journal={Int. J. audiology},
  volume={50},
  number={1},
  pages={2--13},
  year={2011},
  publisher={Taylor \& Francis}
}

@article{roa2025first,
  title={The first Cadenza challenges: using machine learning competitions to improve music for listeners with a hearing loss},
  author={Roa-Dabike, Gerardo and Akeroyd, Michael A and Bannister, Scott and Barker, Jon P and Cox, Trevor J and Fazenda, Bruno and Firth, Jennifer and Graetzer, Simone and Greasley, Alinka and Vos, Rebecca R and others},
  journal={IEEE Open J. Signal Process.},
  year={2025},
  pages={722--734},
  year=2025,
  month=jun,
  volume={6}
}

@article{vanwieringen2008listf,
  title={LIST and LINT: Sentences and numbers for quantifying speech understanding in severely impaired listeners for {Flanders} and the {Netherlands}},
  author={{van Wieringen}, Astrid and Wouters, Jan},
  journal={Int. J. audiology},
  volume={47},
  number={6},
  pages={348--355},
  month=jun,
  year={2008},
  publisher={Taylor \& Francis}
}

@article{jansen2014listm,
  title={Development and validation of the {Leuven} intelligibility sentence test with male speaker (LIST-m)},
  author={Jansen, Sofie and Koning, Raphael and Wouters, Jan and {van Wieringen}, Astrid},
  journal={Int. J. audiology},
  volume={53},
  number={1},
  pages={55--59},
  month=jan,
  year={2014},
  publisher={Taylor \& Francis}
}

@conference{farina_2000_02,
		title={{Simultaneous} {measurement} {of} {impulse response} {and} {distortion} {with} {a} {swept-sine} {technique}},
		author={Farina, Angelo},
		booktitle={Preprints AES 108th Convention},
		publisher={AES},
		month={Feb}, year={2000}, address={Paris, France},
		pages={1--24},
}

@conference{farina_2007_05,
		title={{Advancements} {in} {impulse} {response} {measurements} {by} {sine} {sweeps}},
		author={Farina, Angelo},
		booktitle={Preprints AES 123rd Convention},
		publisher={AES},
		month={May}, year={2007}, address={Vienna, Austria},
		pages={1--21},
}

@article{novak2015synchronized,
  title={Synchronized swept-sine: Theory, application, and implementation},
  author={Novak, Antonin and Simon, Laurent and Lotton, Pierrick},
  journal={J. Audio Eng. Soc.},
  volume={63},
  number={10},
  pages={786--798},
  year={2015},
  month=oct
}

@MISC{ISO-3382-1,
	title={Acoustics - Measurement of room acoustic parameters - Part 1: Performance spaces},
	author={{ISO 3382-1:2009}},
	address={Geneva, Switzerland},
	year={2009},
	howpublished={International Organization for Standardization},
	pages={26}
}

@MISC{ITU-R-BS.1770-4,
	title={Algorithms to measure audio programme loudness and true-peak audio level},
	author={{ITU-R BS.1770-4}},
	address={Geneva, Switzerland},
	year={2015},
	howpublished={International Telecommunication Union},
}

@misc{iosr_matlab_toolbox,
  author = {Hummersone, C. and Pr{\"a}tzlich, T.},
  title = {{IoSR Matlab toolbox}},
  year = {2017},
  howpublished = {GitHub repository},
  url = {https://github.com/IoSR-Surrey/MatlabToolbox}
}

@misc{pyloudnorm,
  author = {C. J. Steinmet and J. D. Reiss},
  title = {{pyloudnorm}},
  year = {2019},
  howpublished = {GitHub repository},
  url = {https://github.com/csteinmetz1/pyloudnorm}
}

@inproceedings{fejgin_brudex_2023,
	title = {{BRUDEX} {database}: {Binaural} {Room} {Impulse} {Responses} with {Uniformly} {Distributed} {External} {Microphones}},
	booktitle = {Proc. Speech Communication; 15th ITG Conf.},
	publisher = {IEEE},
	author = {Fejgin, Daniel and Middelberg, Wiebke and Doclo, Simon},
	month = sep,
	year = {2023},
	address = {Aachen, Germany},
	pages = {126--130},
}

@inproceedings{delebecque_binaurec_2023,
	address = {Helsinki, Finland},
	title = {{BinauRec}: {A} dataset to test the influence of the use of room impulse responses on binaural speech enhancement},
	urldate = {2023-09-11},
	booktitle = {Proc. 31st European Signal Process. Conf. (EUSIPCO '23)},
	publisher = {EURASIP},
	author = {Delebecque, Louis and Serizel, Romain},
	month = sep,
	year = {2023},
	pages = {126--130},
}

@inproceedings{jeub_binaural_2009,
	address = {Santorini, Greece},
	title = {A binaural room impulse response database for the evaluation of dereverberation algorithms},
	isbn = {978-1-4244-3297-4},
	language = {en},
	booktitle = {Proc. 2009 IEEE Int. Conf. Digital Signal Process. (DSP '09)},
	publisher = {IEEE},
	author = {Jeub, Marco and Schafer, Magnus and Vary, Peter},
	month = jul,
	year = {2009},
	pages = {1--5},
}

@article{kayser_database_2009,
	title = {Database of {multichannel} {in}-{ear} and {behind}-the-{ear} {head}-{related} and {binaural} {room} {impulse} {responses}},
	volume = {2009},
	issn = {1687-6180},
	language = {en},
	number = {1},
	journal = {EURASIP J. Adv. Signal Process.},
	author = {Kayser, H. and Ewert, S. D. and Anem{\"u}ller, J. and Rohdenburg, T. and Hohmann, V. and Kollmeier, B.},
	month = dec,
	year = {2009},
	pages = {298605},
}

@article{denk_adapting_2018,
	title = {Adapting {hearing} {devices} to the {individual} {ear} {acoustics}: {database} and {target} {response} {correction} {functions} for {various} {device} {styles}},
	volume = {22},
	issn = {2331-2165},
	language = {en},
	journal = {Trends Hearing},
	author = {Denk, Florian and Ernst, Stephan M. A. and Ewert, Stephan D. and Kollmeier, Birger},
	month = jan,
	year = {2018},
	pages = {2331216518779313--2331216518779313},
}

@misc{maazaoui_romeo-hrtf_nodate,
	title = {Romeo-{HRTF}: {A} {multimicrophone} {Head} {Related} {Transfer} {Functions} {database}},
	language = {en},
	author = {Maazaoui, Mounira and Grenier, Yves},
	url={https://adasp.telecom-paris.fr/resources/2011-03-31-romeo-hrtf},
	month = mar,
	year = {2011},
}

@inproceedings{algazi_cipic_2001,
	address = {New Platz, NY, USA},
	title = {The {CIPIC} {HRTF} database},
	booktitle = {Proc. 2001 IEEE Workshop Appls. Signal Process. Audio Acoust. (WASPAA '01)},
	publisher = {IEEE},
	author = {Algazi, V.R. and Duda, R.O. and Thompson, D.M. and Avendano, C.},
	month=oct,
	year = {2001},
	pages = {99--102},
}

@inproceedings{woods_real-world_2015,
	title = {A real-world recording database for ad hoc microphone arrays},
	address = {New Platz, NY, USA},
	booktitle = {Proc. 2015 IEEE Workshop Appls. Signal Process. Audio Acoust. (WASPAA '15)},
	author = {Woods, William S. and Hadad, Elior and Merks, Ivo and Xu, Buye and Gannot, Sharon and Zhang, Tao},
	month = oct,
	year = {2015},
	publisher = {IEEE},
	pages = {1--5},
}

@misc{noauthor_3d_nodate,
	title = {{3D} {SOUND}: {HRTF} {measurements}},
	author = {Gardner, Bill and Martin, Keith},
	url = {https://www.lim.di.unimi.it/IEEE/BEGAULT/INDEX.HTM},
	month = may,
	year=1994,
}

@inproceedings{lamba_dummy_2024,
	title = {Dummy head {HRTFs} with different head-above-torso orientations},
	language = {en},
	author = {Lamba, Manan and Ohlmann, Kristin and Ewert, Stephan D and Kollmeier, Birger},
	booktitle = {Proc. DAS|DAGA 2024},
	publisher={DEGA e.V.},
	month = mar,
	year = {2024},
	address = {Hannover, Germany},
	pages = {1607--1609}
}

@inproceedings{thiemann_multiple_nodate,
	title = {Multiple {model} {high}-{spatial} {resolution} {HRTF} {measurements}},
	language = {en},
	author = {Thiemann, J and Escher, A},
	booktitle = {Proc. DAS|DAGA 2015},
	publisher={DEGA e.V.},
	month = mar,
	year = {2015},
	address = {Nuremberg, Germany},
	pages = {797--798},
}

@inproceedings{sass_comparison_2010,
	title = {Comparison of recording methods for measurements of individualized head-related transfer functions},
	language = {en},
	author = {Sass, Rebecca and Werner, Stephan and Siegel, Andr{\'e}},
	booktitle={Proc. 2010 Tonmeistertagung-VDT Int. Conv.},
	month=nov,
	year = {2011},
	pages = {721--726},
	address = {Leipzig, Germany},
}

@article{oreinos_measurement_2013,
	title = {Measurement of a {full} {3D} {set} of {HRTFs} for {in}-{ear} and {hearing} {aid} {microphones} on a {Head} and {Torso} {Simulator} ({HATS})},
	volume = {99},
	issn = {16101928},
	language = {en},
	number = {5},
	journal = {Acta Acust. United Ac.},
	author = {Oreinos, Chris and Buchholz, J{\"o}rg M.},
	month = sep,
	year = {2013},
	pages = {836--844},
}

@misc{hettler_acoustic_2024,
	title = {Acoustic feedback tendency in hearing aids for different types and couplings in relation to insertion gain},
	doi = {10.5281/zenodo.10835180},
	language = {eng},
	publisher = {Zenodo},
	author = {Hettler, Fabian and Denk, Florian and J{\"u}rgens, Tim and Husstedt, Hendrik},
	month = mar,
	year = {2024},
}

@article{denk_hearpiece_2020,
	title = {The {Hearpiece} database of individual transfer functions of an openly available in-the-ear earpiece for hearing device research},
	author = {Denk, Florian and Kollmeier, Birger},
	journal = {Acta Acust.},
	volume = {5},
	number = {2},
	pages={2},
	month=dec,
	year = {2021},
}

@article{spriet_combined_2007,
	title = {Combined feedback and noise suppression in hearing aids},
	volume = {15},
	issn = {1558-7924},
	number = {6},
	journal = {IEEE Trans. Audio Speech Lang. Process.},
	author = {Spriet, A. and Rombouts, G. and Moonen, M. and Wouters, J.},
	month = aug,
	year = {2007},
	pages = {1777--1790},
}

@misc{moore_otimp_2019,
	title = {{OTIMP}: {The} {Oticon}-{Imperial} hearing aid impulse response database},
	shorttitle = {{OTIMP}},
	doi = {10.5281/zenodo.5040465},
	language = {eng},
	publisher = {Zenodo},
	author = {Moore, Alastair H and de Haan, Jan Mark and Pedersen, Michael Syskind and Naylor, Patrick A. and Brookes, Mike and Jensen, Jesper},
	month = may,
	year = {2019},
}

@article{sankowsky_on_2015,
	title = {Reciprocal measurement of acoustic feedback paths in hearing aids},
	volume = {138},
	issn = {0001-4966},
	number = {4},
	journal = {J. Acoust. Soc. Amer.},
	author = {Sankowsky-Rothe, Tobias and Blau, Matthias and Schepker, Henning and Doclo, Simon },
	month = oct,
	year = {2015},
	pages = {EL399--EL404}
}

@article{damiano2025trajectorirdatabaseroomacoustic,
      title={The trajectoRIR Database: Room Acoustic Recordings Along a Trajectory of Moving Microphones}, 
      author={Stefano Damiano and Kathleen MacWilliam and Valerio Lorenzoni and Thomas Dietzen and Toon van Waterschoot},
      year={2025},
	  month=dec,
	  journal = {EURASIP J. Audio, Speech, Music Process.},
	  pages={1--45},
}

@INPROCEEDINGS{li_icassp_2026,
  author={Li, Chenda and Wang, Wei and Sach, Marvin and Zhang, Wangyou and Saijo, Kohei and Cornell, Samuele and Fu, Yihui and Ni, Zhaoheng and Fingscheidt, Tim and Watanabe, Shinji and Qian, Yanmin},
  booktitle={Proc. 2026 IEEE Int. Conf. Acoust., Speech, Signal Process. (ICASSP '26)}, 
  title={ICASSP 2026 Urgent Speech Enhancement Challenge}, 
  year={2026},
  pages={21919--21921},
  month=may,
  address={Barcelona, Spain},
 }

@book{naylor_speech_2010,
	author = {Naylor, Patrick A. and Gaubitch, Nikolay Dian.},
	series = {Signals and communication technology},
	title = {Speech dereverberation },
	year = {2010},
	address = {New York},
	booktitle = {Speech dereverberation},
	edition = {1st},
	isbn = {9781849960564},
	language = {eng},
	publisher = {Springer},
}

@book{kates_digital_2008,
	address = {San Diego},
	title = {Digital hearing aids},
	isbn = {978-1-59756-317-8},
	publisher = {Plural Pub.},
	edition = {1st},
	author = {Kates, James M.},
	year = {2008},
}

@article{de_sena_modeling_2015,
	title = {On the modeling of rectangular geometries in room acoustic simulations},
	volume = {23},
	issn = {2329-9304},
	number = {4},
	journal = {IEEE/ACM Trans. Audio Speech Lang. Process.},
	author = {De Sena, E. and Antonello, N. and Moonen, M. and {van Waterschoot}, T.},
	month = apr,
	year = {2015},
	pages = {774--786},
}

@article{brinkmann_round_2019,
	title = {A round robin on room acoustical simulation and auralization},
	volume = {145},
	issn = {0001-4966},
	language = {en},
	number = {4},
	journal = {J. Acoust. Soc. Amer.},
	author = {Brinkmann, Fabian and Asp{\"o}ck, Lukas and Ackermann, David and Lepa, Steffen and Vorl{\"a}nder, Michael and Weinzierl, Stefan},
	month = apr,
	year = {2019},
	pages = {2746--2760},
}

@article{trainor2004development,
  title={Development of a flexible, realistic hearing in noise test environment (R-HINT-E)},
  author={Trainor, Laurel and Sonnadara, Ranil and Wiklund, Karl and Bondy, Jeff and Gupta, Shilpy and Becker, Suzanna and Bruce, Ian C and Haykin, Simon},
  journal={Signal Process.},
  volume={84},
  number={2},
  pages={299--309},
  year={2004},
  month=feb
}

@inproceedings{lydaki_deep_feedback_2025,
	author={Lydaki, Eleftheria and Tan, Zheng-Hua and Jensen, Jesper and Guo, Meng},
	booktitle={Proc. 2025 IEEE Int. Conf. Acoust., Speech, Signal Process. (ICASSP '25)}, 
	title={Deep Feedback Cancellation for Hearing Aids with Improved System Stability and Sound Quality}, 
	month=apr,
	year={2025},
	pages={1--5},
	address={Hyderabad, India},
}

@book{dillon_hearing_2012,
	address = {Sydney, Australia},
	edition = {2nd edition},
	title = {Hearing Aids},
	isbn = {978-1-60406-810-8},
	publisher = {Thieme},
	author = {Dillon, Harvey},
	month = jun,
	year = {2012},
}

\end{document}